\documentclass[journal]{IEEEtran}
\pdfoutput=1 
\usepackage{graphicx}
\usepackage{amsmath}
\usepackage{wrapfig}
\usepackage{multirow}
\usepackage{colortbl}
\usepackage{color}
\usepackage{soul}
\usepackage{enumerate}
\usepackage{subfigure}
\usepackage[table]{xcolor}
\usepackage{pifont}
\usepackage[normalem]{ulem}

\IEEEoverridecommandlockouts
\usepackage{graphicx}
\usepackage{textcomp}
\usepackage{gensymb}
\usepackage{stfloats}
\usepackage{tikz}
\usepackage{caption}
\usepackage{verbatim}
\usetikzlibrary{chains,positioning,shapes.symbols,arrows,shapes,shadows,trees}
\usepackage{standalone}
\usepackage{array}
\usepackage{cite}
\usepackage{enumitem}

\usepackage[font={small}]{caption}
\usepackage{setspace}

\begin{document}
		\bstctlcite{IEEEexample:BSTcontrol}
		\setlength{\parskip}{0pt}
%
\title{Information and Communication Theoretical Understanding and Treatment of Spinal Cord Injuries: State-of-the-art and Research Challenges}

\author{Ozgur B. Akan,~\IEEEmembership{Fellow,~IEEE}, Hamideh Ramezani,~\IEEEmembership{Student Member,~IEEE}, Meltem Civas,~\IEEEmembership{Student Member,~IEEE}, Oktay Cetinkaya,~\IEEEmembership{Member,~IEEE}, Bilgesu A. Bilgin,~\IEEEmembership{Member,~IEEE} and Naveed A. Abbasi,~\IEEEmembership{Student Member,~IEEE}\\ 
	\thanks{O. B. Akan is with the Internet of Everything (IoE) Group, Department of Engineering, University of Cambridge, UK and the Next-generation and Wireless Communications Laboratory (NWCL), College of Engineering, Koc University, Istanbul, Turkey. (e-mail: oba21@cam.ac.uk.) 
	}
	\thanks{H. Ramezani and B. A. Bilgin are with the Internet of Everything (IoE) Group, Electrical Engineering Division, Department of Engineering, University of Cambridge. (e-mails: \{hr404, bab46\}@cam.ac.uk.)}
	\thanks{M. Civas, O. Cetinkaya, and N. A. Abbasi are with the Next-generation and Wireless Communications Laboratory (NWCL), Department of Electrical and Electronics Engineering, Koc University, Istanbul, Turkey (e-mails: \{mcivas16, okcetinkaya13, nabbasi13\}@ku.edu.tr).}
	\thanks{This work was supported in part by ERC project MINERVA (ERC-2013-CoG \#616922), AXA Chair for Internet of Everything and Huawei Graduate Research Scholarship.}}

\maketitle

\begin{abstract}
Among the various key networks in the human body, the nervous system occupies central importance. The debilitating effects of spinal cord injuries (SCI) impact a significant number of people throughout the world, and to date, there is no satisfactory method to treat them. In this paper, we review the major treatment techniques for SCI that include promising solutions based on information and communication technology (ICT) and identify the key characteristics of such systems. We then introduce two novel ICT-based treatment approaches for SCI. The first proposal is based on neural interface systems (NIS) with enhanced feedback, where the external machines are interfaced with the brain and the spinal cord such that the brain signals are directly routed to the limbs for movement. The second proposal relates to the design of self-organizing artificial neurons (ANs) that can be used to replace the injured or dead biological neurons. Apart from SCI treatment, the proposed methods may also be utilized as enabling technologies for neural interface applications by acting as bio-cyber interfaces between the nervous system and machines. Furthermore, under the framework of Internet of Bio-Nano Things (IoBNT), experience gained from SCI treatment techniques can be transferred to nano communication research.
\end{abstract}

\section{Introduction}
The nervous system is one of the most important networks of the body that forms the basis of human intellect and stores our experiences as memories. It is composed of networks of neurons connected through synapses to receive, transmit, and process information in the body. The manipulation of information by the nervous system is still not entirely understood. Thus, an information and communication technology (ICT) framework is required to evaluate the relationships of information exchange between various entities of the nervous system, and then to address the associated problems \cite{akan2016Fundamentals}. 

Communication between the brain and limbs is maintained by the spinal cord, which is comprised of ascending and descending spinal pathways \cite{Waxman_2013}. Each spinal pathway constitutes of neurons having long axons, projecting between the brain and spinal cord. In patients with spinal cord injuries (SCI), advanced amyotrophic lateral sclerosis (ALS), or brain-stem strokes, the brain is functional; however, its signals are not transmitted to the muscles due to interruption in the transmission pathways. 

The social and economic burden of SCI is globally extensive. According to the findings \cite{james2019global}, 0.93 million new cases of SCI occurred in 2016, while prevalent cases were 27.04 million. SCI results in partial loss of motor and/or sensory functions in the incomplete injury. In the case of complete SCI, the patient suffers from the total loss of motor and sensory abilities. Statistics for the United States provided in Table~\ref{Table:SCI_stats} reveal that recovery after injury is very unlikely, and in most cases SCI causes neurological deficits in all limbs and torso, i.e., tetraplegia \cite{SC_facts}. In the less severe form of SCI, paraplegia, the arms are not affected.

\begin{table}[t]
	\def\arraystretch{1.3}
	\footnotesize
	\centering
	\caption{Types of SCI and Corresponding Percentages.}
	\begin{tabular}{|l|c|} \hline
		\rowcolor{gray!10}
		\textbf{SCI Types} & \textbf{Percentage} \\ \hline\hline
		{\emph{Incomplete Tetraplegia}}& $45\%$ \\ \cline{2-2}\hline		
		{\emph{Incomplete Paraplegia}}& $21.3\%$\\ \cline{2-2}\hline			
		{\emph{Complete Paraplegia}}& $20\%$ \\ \cline{2-2}\hline
		{\emph{Complete Tetraplegia}}& $13.3\%$ \\ \cline{2-2}\hline	
		{\emph{Recovery}}& $0.4\%$ \\ \cline{2-2}\hline													
	\end{tabular}
	\label{Table:SCI_stats}
	\vspace{-2mm}
\end{table}

Two main approaches exist for re-establishing the brain-body connection of the patients suffering from SCI, which we group as, biological methods and ICT-based treatment techniques. The latter is the main interest of this paper since it has advantageous aspects in terms of functional outcome, human trials, and safety. The ICT-based treatment techniques of SCI can be achieved by (i) the use of neural interface systems (NIS), and (ii) deployment of artificial neurons (ANs). NIS were first developed for animals \cite{fetz1999real,wessberg2000real,serruya2002brain,taylor2002direct,guo2014encoding,gok2016prediction} and then utilized for human patients to extract control signals from nervous system and operate assistive devices, including computer cursors \cite{kennedy1998restoration,kennedy2000direct,musallam2004cognitive,hochberg2006neuronal,simeral2011neural} and robotic prostheses \cite{hochberg2012reach,collinger2013high,velliste2008cortical,soekadar2016hybrid}. More importantly, they were also used to apply electrical stimulations to the nervous system or muscles for restoring motor capabilities \cite{moritz2008direct,pohlmeyer2009toward,ethier2012restoration,zimmermann2014closed,capogrosso2016brain,bouton2016restoring}, which can induce partial neurological recovery \cite{harkema2011effect,donati2016long}. The deployment of ANs, on the other hand, targets to replace injured or dead biological neurons. 
This is a relatively recent direction, and there is still no major work that can qualify itself as a fully-functional AN.

In this paper, we first discuss the mechanisms of SCI to give some insight on these injuries. We then present the existing treatment techniques including biological, NIS, and AN-based approaches. Specifically, for the NIS and the AN, where the ICT techniques may benefit from the existing designs, we also identify the key characteristics of a successful treatment technique. Based on the preceding findings and discussions, we then present two potential solutions that may be able to treat SCI with their respective methodologies. The first approach is based on NIS, where we identify key areas for improvement in the existing setups. We point out important areas, such as feedback, that are yet unexplored. The second proposal is for the development of ANs based on nanomachines that act as a bridge to deliver neural signals across the injury site. The proposed AN will incorporate plasticity along with self-organization, and will thus be able to adapt in the injury area to make new nervous connections. Additionally, the machines are proposed to include energy harvesting (EH) so that they are truly independent in nature. Apart from SCI treatment, the proposed schemes can be used as bio-cyber interfaces in a variety of applications and also motivate for Internet of Bio-Nano Things (IoBNT) and intra-body nanonetworks applications.

The rest of this paper is organized as follows. In Section~II, we present an overview of nervous communication, the mechanisms for SCI, and some key treatment approaches. The existing ICT-based treatment techniques for SCI are detailed in Section III. Section~IV introduces our proposed methods, whereas Section~V provides some future directions in SCI treatment as well as other potential applications for the SCI treatment techniques. Finally, Section~VI concludes the paper.

\section{Spinal Cord Injury Treatment Techniques}
In this section, we present an overview of nervous system communication, how it is disrupted by SCI during various phases of the injuries, the existing biological treatment techniques for SCI, and the key differences between the biological and ICT-based treatment techniques.

\subsection{Neuro-spike Communication} 
Communication in the ultra large-scale nervous network, known as neuro-spike communication, takes place through electrical or chemical synapses with the latter occurring more frequently \cite{balevi2013physical}. Information encoded in electrical impulses is transmitted to the post-synaptic neuron using three steps. \textit{(i) Axonal transmission}: the electrical impulses pass through the axon of the pre-synaptic neuron until they reach the pre-synaptic terminals. \textit{(ii) Synaptic propagation}: arrival of an electrical impulse to the pre-synaptic terminal initiates the release of neurotransmitter into the synaptic cleft, i.e., the gap between input and output neuron. The neurotransmitters then diffuse though the synaptic cleft to reach the post-synaptic neuron and bind to the receptors located on its membrane. This changes the post-synaptic membrane potential. \textit{(iii) Spike generation}: Depending on the released neurotransmitters, changes in post-synaptic membrane potential can be both excitatory, i.e., a temporary increase in the postsynaptic membrane potential caused by the flow of positively charged ions into the postsynaptic cell, or inhibitory, which makes a postsynaptic neuron less likely to generate an action potential.
Postsynaptic potentials are subject to spatial and temporal summation. Spatial summation
occurs when a neuron is receiving input from multiple synapses that are near each other. Temporal summation occurs when a neuron receives inputs that are close together in time, i.e., the next change in membrane potential happens before the previous change is vanished. If the total change in the post-synaptic membrane potential due to all excitatory and inhibitory synapses is greater than the spiking threshold, the post-synaptic neuron fires an electrical impulse.

Several studies exist in the literature on modeling and analyzing the performance of neuro-spike communication \cite{malak2014communication,balevi2013physical,malak2013communication,ramezani2017communication,Ramezani2015,Ramezani2017Importance,Ramezani2017Rate,ramezani2017information,khan2017diffusion} and the nervous nanonetwork \cite{abbasi2015queueing,abbasi2018controlled} considering the functionality of healthy neurons. Any disruption in the functionality of neurons, which can result from diseases or injuries, adversely affects the performance of this nanoscale communication network. 

\begin{figure}[t]
	\centering
	\includegraphics[width=9cm]{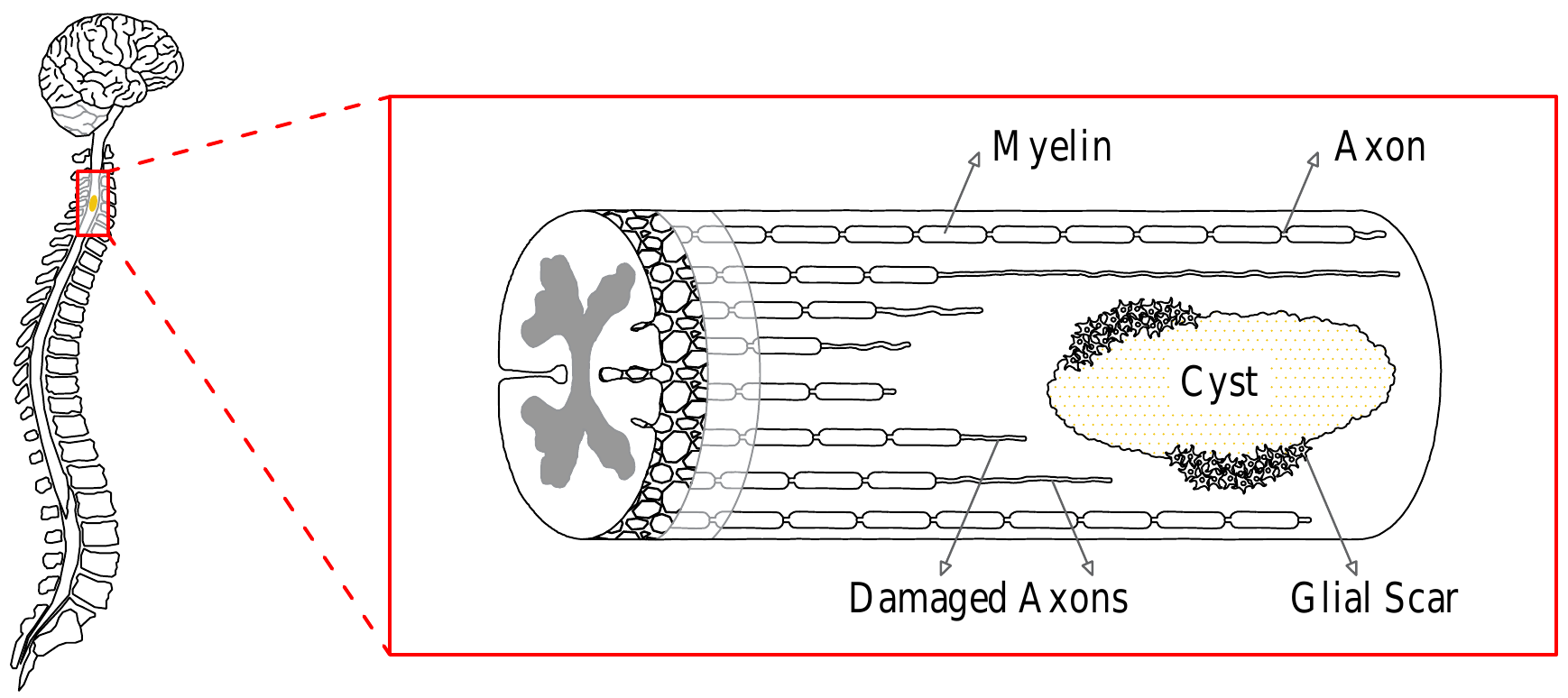}
	\caption{SCI and the formation of glial scars. The demyelination, i.e., the loss of myelin sheath, occurring due to neuroinflammation significantly affects the axonal propagation, which might get disrupted in severe cases.}
	\label{fig:damaged_SC}
\end{figure}

\subsection{Spinal Cord Circuitry}
The nervous system comprises two main systems: the peripheral and central nervous networks. A part of the peripheral nervous network (PNN) called somatic nervous network (SNN) is related to the control of voluntary body movements. Voluntary body movements are controlled by conveying sensory inputs to central nervous network (CNN) via sensory nerves and motor outputs from CNN to the skeletal muscles by motor nerves. Motor centers in the brain generate motor outputs which are integrated by spinal cord with the sensory inputs. The response generated by the spinal cord is transmitted to the muscles \cite{purves2004neuroscience}. Spinal cord circuitry consists of several complex networks, which can be divided as motor network and sensory network from the communication theory perspective.

\begin{figure*}[!t]
	\centering
	\subfigure[Diagram showing a spinal cord motor network.]{\label{fig:ra}\includegraphics[width=0.4\textwidth]{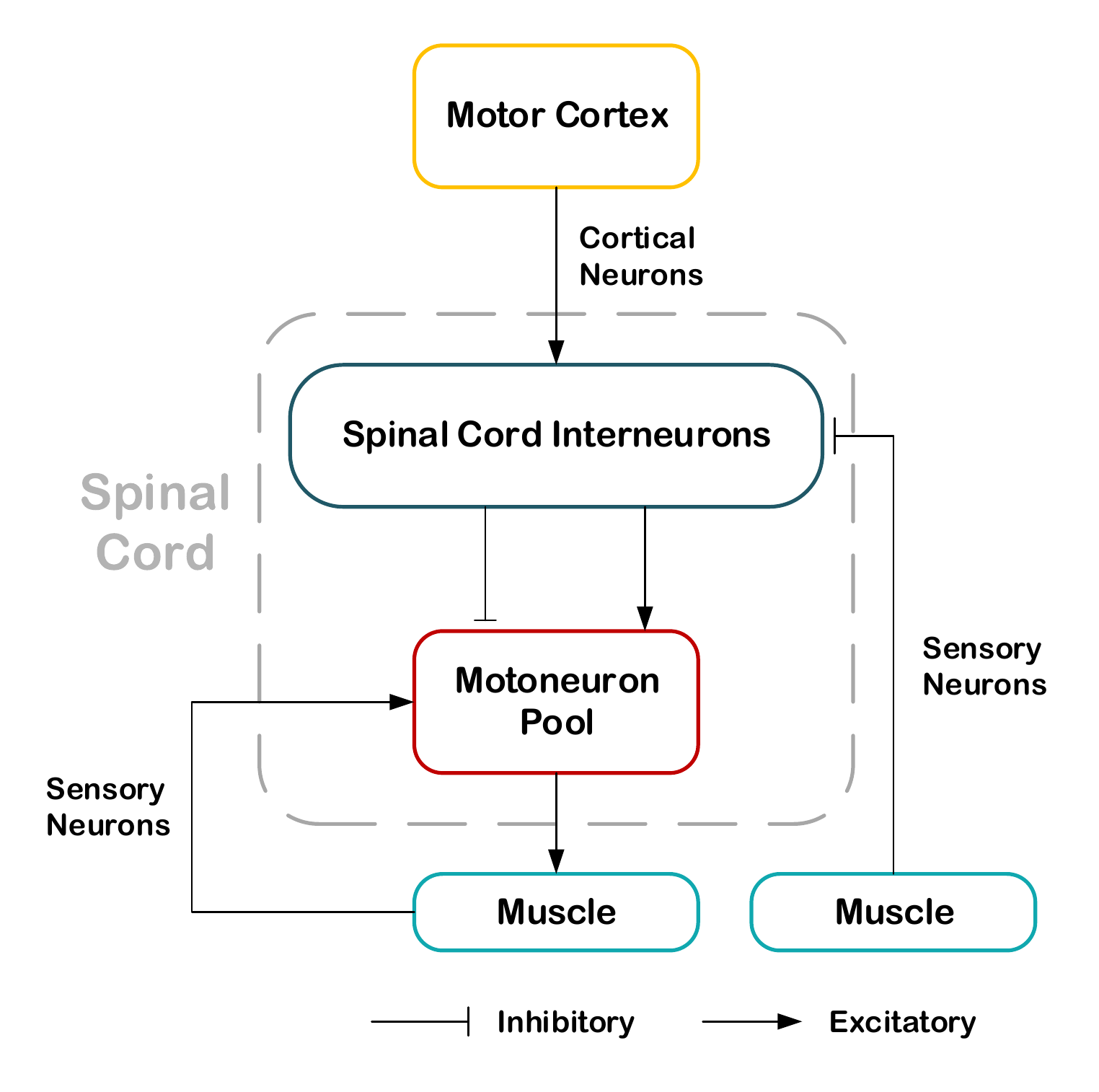}}
	\subfigure[A spinal cord interneuronal network.]{\label{fig:rb}\includegraphics[width=0.4\textwidth]{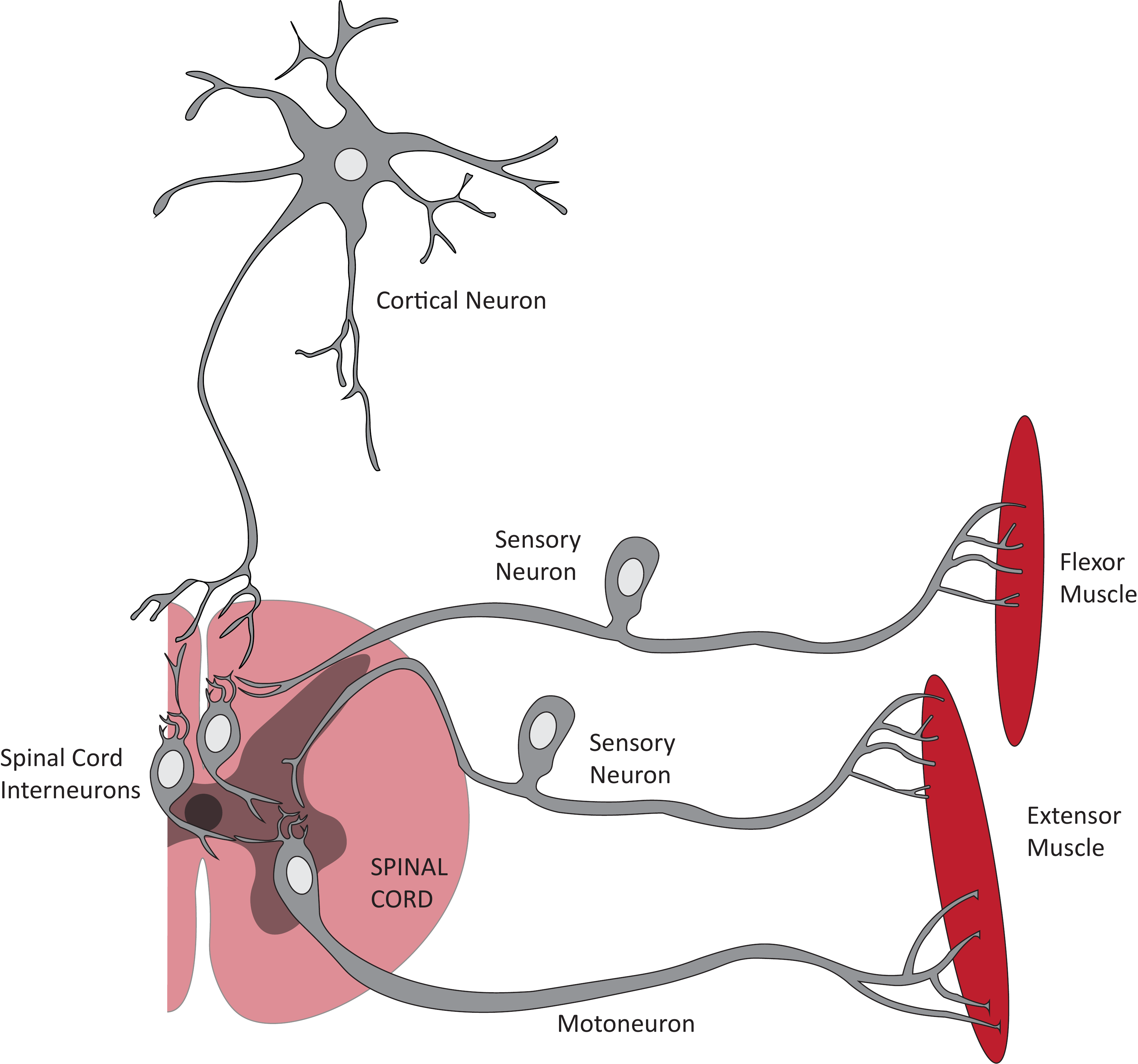}}
	\caption{Diagram illustrating spinal cord sensory-motor networks \cite{civas2020rate}.}
	\label{fig:revision}
\end{figure*}

Motor network performs the downlink transmission of the motor commands from the brain to the muscles. As depicted in Fig. \ref{fig:ra}, cortical neurons transmit motor inputs to the spinal cord interneurons, which integrate motor inputs conveyed from the brain and the sensory inputs transmitted from the sensory organs in the muscles. In primates, cortical neurons directly project to motoneurons \cite{lemon2008descending}. A motoneuron pool in the spinal cord is the final destination for the sensory-motor inputs controlling a specific muscle. Thus, motoneurons are important for both fine and gross motor control. Moreover, they connect to sensory neurons via the spinal
 interneurons in reflex activity.
 
 Sensory network controls the uplink transmission of sensory information from the skeletal muscles to the brain by using sensory pathways comprising sensory nerves. Sensory modalities 
 from the body are detected by afferent fibers whose cell bodies
 reside in dorsal root ganglion. These primary afferents mostly
 terminate in the spinal cord, where sensory information is first
 processed and synapse with interneurons or projection neurons
 forming tracts. Projections that course through ascending
 pathways convey messages to higher targets in the central nervous network including brainstem and somatosensory cortex. 
 
 In the case of damage to the cortical neurons with long axons projecting to the spinal cord as a result of SCI, the motor inputs for the voluntary movements may not be delivered reliably to the spinal cord, where the spinal cord interneurons integrate the motor inputs from the brain and the sensory outputs from the muscles. To activate the spinal cord interneuron circuit, the motor commands read from the motor cortex area related to fine and grass motor movements can be delivered artificially to the spinal cord interneuron circuits. Considering SCI resulting in a damage to the sensory network, the sensory signal can be read from the dorsal root ganglion, which contains the nerve bundles of first order sensory neurons. Then, the sensory signal obtained can be delivered to the related parts of the somatosensory cortex. To differentiate the sensory signals from each other signal processing techniques can be employed.

\color{black}
\subsection{Injury Mechanisms} 
Neurons inflicted by spinal trauma experience the effects of an injury in three consecutive phases, namely, primary, secondary, and chronic injuries described below.
\begin{itemize}[leftmargin=*]
	\item \textit{Primary injury:} This refers to the condition within minutes after the injury happens. The injury immediately disrupts axonal pathways and damages cell membranes depending on the physical trauma \cite{McDonald_2002}. The ischaemia, i.e., toxic chemical release from damaged membranes, initiates the secondary injury \cite{McDonald_2002}. Many active neural connections, and thus neuro-spike communications, are affected by this extracellular toxicity.
	
	\item \textit{Secondary injury:} Following hours to weeks after the injury, SCI develops a condition called secondary injury \cite{Silva_2014}. In this phase, deteriorated toxic environment in the spinal cord causes death of neurons as well as glial cells.  \cite{11_papastefanakimatsas_2015}. 
	Glial cells produce myelin sheath, which acts as an insulator and improves the rate of information transmission along the axon. The loss of these special cells significantly affects axonal propagation due to loss of myelin sheath, namely demyelination \cite{Oyinbo_2011}. Moreover, axonal regeneration is blocked by glial scar formation, which is a tissue barrier formed in the injury site, as depicted in Fig.~\ref{fig:damaged_SC}. In severe cases, axonal transmission is blocked due to demyelination and disruption of the brain-spinal cord communication \cite{Oyinbo_2011}.   
	
	\item \textit{Chronic injury:} This	phase covers the days to years following the injury \cite{Silva_2014}. In this phase, injured axons experience a special kind of degeneration, called Wallerian degeneration, that disconnects the distal parts of the axons from their cell bodies while the glial scar continues to grow \cite{5_kabu_2015}. Wallerian degeneration consists of several phases, in which axonal death occurs and axon debris is cleared by the glial cells \cite{llobet2019axon}. However, the proximal parts of the motor fibers are preserved and still functioning. The time course of the degeneration of cell bodies is very slow compared to that of axonal degeneration. To illustrate, even years after the injury, cell bodies can be preserved \cite{prasad2012can}. This makes the recording motor signals from the spinal cord descending tracts possible \cite{prasad2012can}. Any damage to the long axons of descending spinal cord tract conveying information of motor actions will damage to the spinal cord motor network. However, the circuitry below the injury will remain intact.
\end{itemize} 

The overall result of SCI mechanisms is the disruption of axonal pathways located within the spinal cord, which results in either the loss of sensory information transmission to the brain or motor control signals to the muscles or both at the same time \cite{McDonald_2002}. Hence, re-establishing the brain-spinal cord communication is necessary to maintain the sensory and motor functions of the nervous nanonetwork. As discussed earlier, current treatment approaches fall in two main categories, namely the biological and the ICT-based treatment techniques. In the following, we discuss the major techniques that belong to these categories and the key differences between them.

\subsection{Biological Treatment Techniques}
Following the initial disruption of SCI, damage continues to spread in the spinal cord in the secondary and the chronic phases. Despite to ongoing research and clinical trials on preventing the effects of pathological results of SCI, a gold standard treatment of SCI is not known \cite{kim2017spinal}. Preventive and treatment strategies can be classified as biological and ICT-based approaches. Biological treatment approaches aims to protect surviving neurons from the toxic post-injury environment, and to recover synaptic connections and healthy functionality by promoting regeneration. On the contrary, ICT-based treatment techniques directly address the neural communication problem by either bypassing the injury site via external communication interfaces or by replacing the injured neurons by artificially designed neurons. 

Next, we outline the existing fundamental biological treatment approaches and the challenges faced regarding their implementation. Furthermore, we discuss how these challenges can be addressed by ICT-based treatment techniques. 

Research efforts regarding biological treatments mainly focus on the following directions: 

\vspace{2mm}
\subsubsection{Neuroprotection}
Delivering drugs onto the injury site is a possible method of neuroprotection. Recent research in this area concentrates on drug delivery mediated by nanomaterials (e.g., nanowires, nanoparticles, and carbon nanofibers), which have several advantages over conventional drug delivery methods, such as providing targeted delivery and reaching injury site by crossing the blood-spinal cord barrier \cite{tyler2013nanomedicine}. 
The challenges in this direction include the following: 

\begin{itemize}[leftmargin=*]
	\item \textit{Application:} Devastating effects of secondary injury mechanisms spread fast with the cascade of biological complex events. Preventing this spread means intervening many biological processes in the spinal cord with presumably number of different neuroprotective agents. Thus, identifying the effective drugs and their administrations in a limited time period poses challenges. Moreover, there are many issues regarding drug carrier design \cite{tyler2013nanomedicine}.      
	\item \textit{Validation:} Several drugs are tested on animals and \textit{in vitro} environments for different injury scenarios. Identifying effective drugs and methods among many alternatives and verifying the efficacy by human trials still pose a challenge. 
	\item \textit{Functional outcome:} Although several neuroprotective drugs are reported to be resulting in some functional recovery \cite{faccendini2017nanofiber}, functional outcome is limited because complete recovery also necessitates the reformation of spinal cord circuitry. 
\end{itemize}

\vspace{2mm}
\subsubsection{Neural Regeneration}
Delivery of neuro-regenerative drugs, cell-based therapy and tissue engineering aim to promote neural regeneration. 
In drug delivery, growth factors, such as neurotrophins, help in improvement of nerve regeneration, synaptic transmission, and plasticity \cite{willerth2007approaches}. Cell therapy efforts mostly focus on cell transplantation and grafting. Transplantation of stem cells is widely investigated as a cell-based therapy \cite{lu2012long,nakajima2012transplantation, sharp2010human} since stem cells have differentiation capabilities that allow them to change into neurons and glia. In addition, a variety of cells are used in studies, such as Schwann cells, that are in-charge of producing myelin sheaths in peripheral nervous system and capable of improving remyelination as well as axonal regrowth \cite{kanno2014combination,11_pearse_2007}. Tissue engineering aims to promote regeneration of injured nerves by means of materials that can mimic natural scaffolding, such as hydrogels, nanofiber scaffolds, and hybrid applications. Hydrogels are biocompatible networks of polymers that can improve the environment for neural regrowth \cite{5_kabu_2015,krsko2009length,mahoney2006three}. Nanofiber scaffolds not only provide supportive scaffolding for damaged cells but also deliver drugs and support transplanted cells \cite{liu2013self,guo2007reknitting,tysseling2008self}. Various combinations of hydrogels, nanofibers, and drugs are also studied for recovery of SCI as shown by \cite{chen2015repair,milbreta2016three,nguyen2017three}. Fundamental challenges in this direction include the following:  

\begin{itemize}[leftmargin=*]
	\item \textit{Functional outcome:} Although the above-mentioned studies report some improvement regarding neural regeneration, and recent research show several initial results regarding functional neural network formation \cite{gautam2017engineering, timashev20163d,demarse2016feed,gladkov2017design}, 
	directing regenerating neurons to form functional circuitry is still challenging \cite{assinck2017cell}. 
	\item \textit{Human Trials:} Safety and efficiency of medicine treatment and cell therapy on humans are still controversial \cite{vismara2017current, assinck2017cell}. Regarding tissue engineering, functionally integrating biocompatible materials to the tissues and the feasibility of human trials pose potential challenges \cite{chen2016advancing}. 
\end{itemize}

\subsection{Biological versus ICT-based Treatments}
Since the ICT-based treatment approaches do not have protective or supportive aims, they directly focus on communication problems between the brain and the spinal cord. Several initial studies have provided promising results for these techniques with regard to safety and the functional outcome. In this respect, some advantages over biological treatment approaches are stated below: 

\begin{itemize}[leftmargin=*]
	\item \textit{Functional outcome:} As processing, stimulation, and communication capabilities of the external interfaces advance with the developments in ICT, restoration in the movements of a subject with SCI can be observed as proved by a recent study \cite{capogrosso2016brain}.  
	\item \textit{Human trials:} In addition to animal experiments, in studies, such as \cite{bouton2016restoring}, promising improvements in terms of motor impairment in human patients were observed. 
	\item \textit{Safety:} Although the long term safety of invasive recording devices in NIS is still not proven, there are less invasive alternatives with low risk neural interfaces, such electroencephalography (EEG)- \cite{rajangam2016wireless} and electrocorticography (ECoG)-based systems. 
\end{itemize}

\section{Existing ICT-based Treatment Techniques}
In this section, we present prevalent ICT-based treatment techniques reported in the literature as well as future directions that may be possible in the NIS and AN frameworks. Two techniques contrast with each other on the scale of their approaches with NIS focusing on macro-scale solutions and AN employing a micro/nano-scale approach.    
\subsection{Neural Interface Systems}
\begin{figure*}[t]
	\centering
	\hspace{-3mm}
	\includegraphics[width=18.5cm]{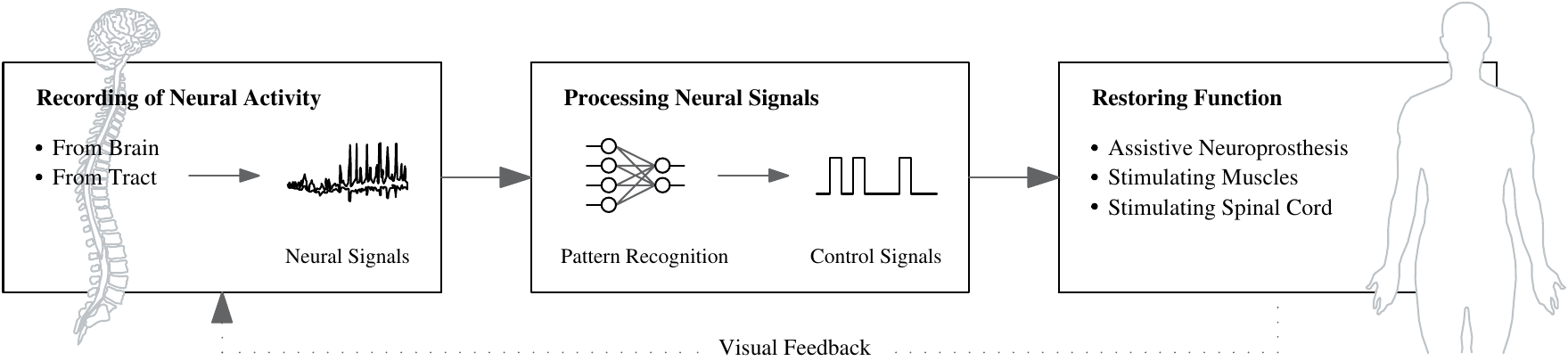}
	\caption{Basic structure of an NIS, which restores the motor function by using the nervous signals directly captured from the brain or just before the injured part of the spinal cord.}
	\label{fig:NIS_Block}
\end{figure*}

While the aforementioned biological techniques aim to improve signal transmission problem through the injured spinal cord, NIS utilize nervous signals captured directly from the brain or from the spinal cord before the injured part to restore the motor function. 
Fig.~\ref{fig:NIS_Block} depicts the major components of a typical NIS, details of each are elaborated below.

\subsubsection{Nervous recordings}
As mentioned, neural signals can be either recorded from the brain \cite{boraud2002single,wolpaw2002brain,kennedy1998restoration,capogrosso2016brain,donati2016long,soekadar2016hybrid,gangulyintroduction} or the spinal cord \cite{guo2014encoding}. While the former is more common in literature, the latter can potentially provide signals with more information from just a small anatomical region. The recording region is one of important factors that needed to be selected and optimized according to the purpose of NIS. The recording region of the existing experiments in the NISs literature are tabulated in Table~\ref{table:recording}.

\begin{table*}[t]
	\centering
	\caption{Nervous recordings in the NIS literature.}
	\resizebox{0.85\textwidth}{!}{
	\begin{tabular}{p{4cm} p{6cm} p{3.9cm} l l} \hline
		\rowcolor{gray!10}
		\bf{Purpose} & \bf{Recording region} &\bf{Recording} &\bf{Subject} &\bf{Ref.}\\ \hline 
		Self modulation of sensorimotor EEG rhythms, related to the imagination of limb movements & Scalp potentials from 59 positions & Electroencephalography (EEG) & Human &\cite{cincotti2008high}\\ \hline		  
		Advanced wheelchair control & Scalp potentials from 32 positions & EEG & Human &\cite{craig2007adaptive}\\ \hline		  
		Control hand exoskeleton& EEG from 5 positions and electro-oculography recordings (EOG) & EEG and EOG & Human &\cite{soekadar2016hybrid}\\ \hline		  
		1D cursor movement & Electrodes placed subdurally on the cortical surface, in some subjects, electrode coverage included sensorimotor or speech cortex areas & Electrocorticography (ECoG) & Human &\cite{leuthardt2006electrocorticography}\\ \hline		  
		Control a 1D cursor & Left frontal-parietal-temporal region including parts of
		sensorimotor cortex  & ECoG & Human &\cite{leuthardt2004brain}\\ \hline		  
		Decoding two-dimensional movement trajectories & Fronto-parietal-temporal
		region including parts of sensorimotor cortex & ECoG & Human &\cite{schalk2007decoding}\\ \hline		  
		Decoding motor intentions & Hand area of
		the ipsilesional primary motor cortex, premotor and somatosensory areas & ECoG & Human &\cite{spuler2014decoding}\\ \hline		  
		Control of 3D cursor movement& Hand and arm area of the left sensorimotor cortex & ECoG & Human &\cite{wang2013electrocorticographic}\\ \hline		 
		Control individual finger movements in real time & sensorimotor regions & ECoG & Human &\cite{hotson2016individual}\\ \hline		  		
		Reaching and grasping & Dorsal premotor cortex (PMd), supplementary motor area (SMA), the primary motor cortex (M1), primary somatosensory cortex (S1), the medial intraparietal area (MIP) of the posterior parietal cortex (PP) & Single unit & Monkey &\cite{carmena2003learning}\\ \hline
		One-dimensional movements of a robot arm & PMd, M1, PP  & Single unit & Monkey &\cite{wessberg2000real}\\ \hline
		A continuous reaching task & M1 & Local field potentials & Monkey &\cite{flint2013long}\\ \hline		  
		Control a 2D cursor & M1, PMD & Local field potentials & Monkey &\cite{so2014subject}\\ \hline		  
		Long-term control of cursor trajectory and click& Arm area of motor cortex & Single unit and local field potentials & Human &\cite{simeral2011neural}\\ \hline
		Restoring weight-bearing locomotion of the paralyzed leg & Leg area of motor cortex & Spiking activity of neuronal ensembles & Monkey &\cite{capogrosso2016brain}\\ \hline				   
		Encoding of forelimb forces & Corticospinal tract (CST) & Recording from spinal tract & Rat&\cite{guo2014encoding}\\ \hline		  
		Prediction of forelimb muscle EMGs & CST & Recording from spinal tract & Rat &\cite{gok2016prediction}\\ \hline		  	
	\end{tabular}}
	\label{table:recording}
\end{table*}

\color{black}The key factors in selecting a recording technique in an NIS include efficacy, safety (particularly with regards to surgery), reliability, and longevity. Furthermore, ease of NIS training, support, cost, and availability are also important factors. In the following, we review signal recording techniques from both of the signal sources, and discuss their advantages as well as limitations.

\paragraph{Recording from the brain}
Brain signals can be recorded via non-invasive techniques, such as EEG, magnetoencephalography (MEG) or functional magnetic-resonance imaging (fMRI). These techniques have low spatio-temporal resolution, low signal-to-noise ratio (SNR), and poor sensitivity to high-frequency changes. Hence, they are insignificant for investigating the short-lived spatio-temporal dynamics of many brain processes \cite{hill2012recording}. However, the feasibility of these techniques in detecting motor intent to some extent is demonstrated in literature, such as \cite{soekadar2016hybrid,donati2016long}, where robotic exoskeletons are operated with use of control signals extracted by EEG and electro-oculography recordings. 

In contrast, recordings with exceptionally high SNR, lower sensitivity to artifacts than EEG, and high spatio-temporal resolution can be achieved by invasive recording techniques such as ECoG and intracortically electrode placements \cite{hill2012recording}. In ECoG, electrodes are implanted subdurally on the surface of the brain; hence, they are able to record the activity of groups of neurons and provide movement-related field potentials. These recordings are robust over long periods \cite{schalk2011brain}, and higher spatio-temporal resolutions can be achieved by utilizing the micro-ECoG grids \cite{wang2009human}. Furthermore, the ECoG signal is used for offline decoding of seven degrees of freedom for arm movements in monkeys and online decoding for individual finger movements in human subjects \cite{chao2010long, hotson2016individual}. This demonstrates that it can provide enough independent control signals for simple loco-motor task. 

However, the implantation of electrodes in the cortex is required to directly record neuronal action or local field potentials from the brain and to increase the spatial resolution of the recordings. This implantation yields to significant clinical risks as a result of infection or the formation of scar tissues. Hence, the electrodes need to be placed on a very small cross sectional area to minimize the damage to the tissue. Moreover, this recording technique requires sophisticated signal processing and computationally intensive algorithms to interpret the neural activity and reliably separate signals of different neurons \cite{cheung2007implantable}. Although point-and-click control is shown to be successfully achieved $1000$ days after implantation of a $100$-electrode array in an individual with tetraplegia \cite{simeral2011neural}, the long-term stability of the implanted electrodes need to be studied in more depth. In particular, intracortical recordings face short-term recording failures due to implant micro-motion, mechanical mismatch of the device and tissue, foreign-body response, and formation of glial scar tissue that interfere with signal transmission \cite{gunasekera2015intracortical}.

\paragraph{Recording from spinal tracts}
Although recording electrical signals from the brain is largely studied in the literature, many technical challenges still exist in the use of microelectrodes to reach a stable recording of individual cell activities. The most important one is the formation of a layer of activated astrocytes around the recording electrode that makes long-term recording of single spikes very difficult. Hence, recording motor control signals through cross-sectional area of corticospinal tract is investigated in \cite{guo2014encoding,gok2016prediction}. The main drawbacks of using this recording source for NIS can be listed as the following:

\begin{itemize}[leftmargin=*]
	\item Designing a mechanically stable array of electrodes is challenging.
	\item The range of signal capture by an electrode array is limited, and thus the number of neural sources are far more than the array recording electrode contacts.  
	\item The control signal cannot be received in corticospinal tract for patients with tetraplegia or ALS.
\end{itemize}

\vspace{2mm}
\begin{table}[t]
	\def\arraystretch{1.3}
	\footnotesize
	\centering
	\caption{A list of methods used in a pattern recognition system.}
	\begin{tabular}{|l|l|c|}
		\hline
		\rowcolor{gray!10}
		\textbf{Process} & \textbf{Method} & \textbf{Ref.}\\ \hline\hline
		&AutoRegressive component&\cite{krusienski2006evaluation} \\ \cline{2-3}
		&Wavelet transform&\cite{cleophas2013machine} \\ \cline{2-3}
		&Common spatial pattern&\cite{alpaydin2004introduction,biship2007pattern} \\ \cline{2-3}
		\multirow{-4}{*}{\emph{Feature extraction}}& Matched filtering &\cite{brunner2010improved} \\ \cline{2-3}\hline		
		&Principal component analysis  &\cite{biship2007pattern,marsland2011machine}\\ \cline{2-3}
		\multirow{-2}{*}{\emph{Dimension reduction}}&Independent component analysis &\cite{biship2007pattern,marsland2011machine} \\ \cline{2-3}\hline			
		&Genetic algorithm & \cite{marsland2011machine}\\ \cline{2-3}
		\multirow{-2}{*}{\emph{Feature selection}}&Sequential selection &\cite{nicolas2012brain} \\ \cline{2-3}\hline
		&Linear discriminant analysis  &\cite{biship2007pattern,marsland2011machine}\\ \cline{2-3}
		& Support vector machine &\cite{biship2007pattern,marsland2011machine}\\ \cline{2-3}
		&Bayesian statistical classifier  &\cite{alpaydin2004introduction,barber2012bayesian}\\ \cline{2-3}																					
		&K-nearest neighbor classifier  &\cite{biship2007pattern,barber2012bayesian} \\ \cline{2-3}
		\multirow{-5}{*}{\emph{Classification}}&Artificial neural network &\cite{goodfellow2016deep} \\ \cline{2-3}\hline							
	\end{tabular}
	\label{Table:Pattern_Recognition}
\end{table}

\subsubsection{Signal Processing}
Real-time processing of the recorded neural data needs to be done for deriving the intended task by the patient. This process is seen as a pattern recognition system that contains three main parts: (i) feature extraction; (ii) dimension reduction and feature selection; (iii) classification or regression. First, the data is processed to derive a set of features. Feature sets extract the discriminating information that represent a dataset. Since the brain signals are the combination of several simultaneous sources and noise, extraction of an appropriate feature set is a challenging task. The data can contain undesirable components, i.e., artifacts, which need to be removed to improve the performance of NIS. Moreover, not all of the recorded signals through multiple channels are relevant for understanding the phenomena of interest. Hence, dimension reduction and feature selection methods are utilized to remove irrelevant and redundant information. 
In Table~\ref{Table:Pattern_Recognition}, a list of methods available for each step of a pattern recognition system is provided \cite{nicolas2012brain}.

Apart from the aforementioned methods, a deep neural network (DNN), which can be trained by a training set of recordings, can also be utilized to extract the task intended by a patient. The advantage of using a DNN is that it extracts the features by itself, which is a complex task as the recorded neural data are high-dimensional and not well-known. Furthermore, the DNN predicts the non-linear systems more accurately \cite{lecun2015deep}.   

After extracting the intended task from neural recordings, control signals must be derived to restore or replace natural function of a paralyzed or lost limb. In case of spinal cord stimulations for restoring the functionality of the injured spinal cord, these control signals are the parameters needed as the input of the spinal cord stimulator. In \cite{capogrosso2016brain}, the extensor and flexor hotspots corresponding to different brain signals are identified in intact monkeys, then the monkeys are paralyzed. Since such training data is not available in paralyzed patients, experiments on healthy subjects are needed to detect whether these hotspots are almost the same in different subjects. Even if these hotspots are placed in different locations for each subject, the epidural stimulation method of \cite{harkema2011effect} together with a machine learning (ML) algorithm can be used to gather the recorded data from healthy subjects and choose stimulation parameters \cite{harkema2011effect}.

\subsubsection{Restoration of Function}
The extracted control signals from the previous steps are then utilized to restore or replace natural function of a paralyzed or lost limb. Such NIS include brain-computer/machine/spine interfaces \cite{serruya2002brain,capogrosso2016brain,soekadar2016hybrid,taylor2002direct,kennedy2000direct,musallam2004cognitive,hochberg2006neuronal,simeral2011neural,hochberg2012reach,collinger2013high,velliste2008cortical,moritz2008direct,pohlmeyer2009toward,ethier2012restoration,zimmermann2014closed,bouton2016restoring,donati2016long,nicolas2012brain}, or spinal cord-computer interfaces \cite{guo2014encoding,gok2016prediction} that are described below:

\paragraph{Operating assistive neuroprosthesis}
Real-time control of robotic prosthesis and exoskeleton finds its applications in practical rehabilitation by replacing the lost motor function to support of daily actions. Studies on using intra-cortical recordings of the brain to control robotic prosthesis \cite{hochberg2012reach,collinger2013high,velliste2008cortical} demonstrate that paralyzed patients may recreate multidimensional control of complex devices even years after the injury. This is utilized in \cite{soekadar2016hybrid} to control a hand exoskeleton that helps patients in restoration of independent daily life activities. Moreover, with sufficient trainings and use of EEG-controlled robotic actuators, it is shown that patients achieve the ability to perform voluntary motor control in key muscles below the spinal cord injury level \cite{donati2016long}. However, using assistive neuroprosthesis is not the focus of this study since it does not provide a solution for bypassing the injured part of the spinal cord to restore the control over muscles.

\paragraph{Direct stimulation of muscles}
Direct stimulation of muscles by signals extracted from intra-cortical recordings was first studied on monkeys with a transiently paralyzed arm \cite{moritz2008direct,ethier2012restoration,pohlmeyer2009toward}, and then applied to a paralyzed human \cite{bouton2016restoring}. In both studies, the exact control signals for each of the muscles required in an activation are extracted and then applied to stimulate the muscles. Results suggest that intra-cortical recordings can be linked to muscle activation in real-time, enabling the control of muscles using the activity of neurons in the motor cortex. This process essentially bypasses the spinal cord and restores the voluntary control of the paralyzed muscles. However, electrical stimulation of muscles may cause fatigue, and long term stimulation by electrodes can damage muscles.

\paragraph{Spinal cord stimulation}
Epidural spinal cord stimulation and intraspinal microstimulation are two existing methods in the literature for inserting motor control signals back to the motor circuitries. In the first one, the stimulating electrodes are placed in the epidural space of the spinal cord \cite{harkema2011effect}, while the electrodes are implanted within the ventral gray matter \cite{dube2014fatigue,grahn2015wireless} in the second method.

\begin{figure*}[!t]
	\centering
	\subfigure[Yellow light activates the opsin called halorodopsin turning the neuron off.]{\label{fig:Opto_off}\includegraphics[width=75mm]{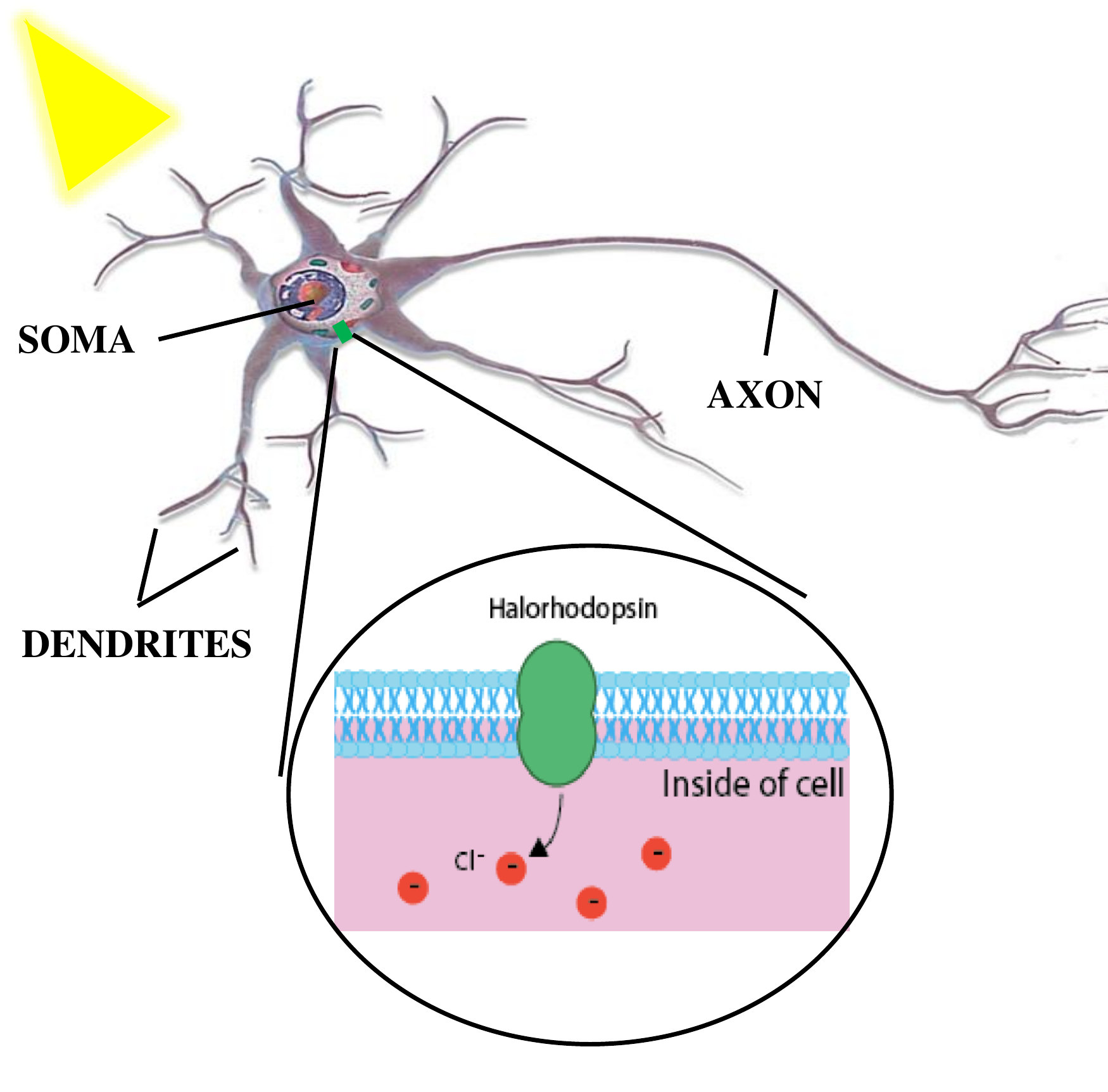}}
	\hspace{5mm}
	\subfigure[Blue light activates the opsin called channelrhodopsin turning the neuron on.]{\label{fig:Opto_on}\includegraphics[width=75mm]{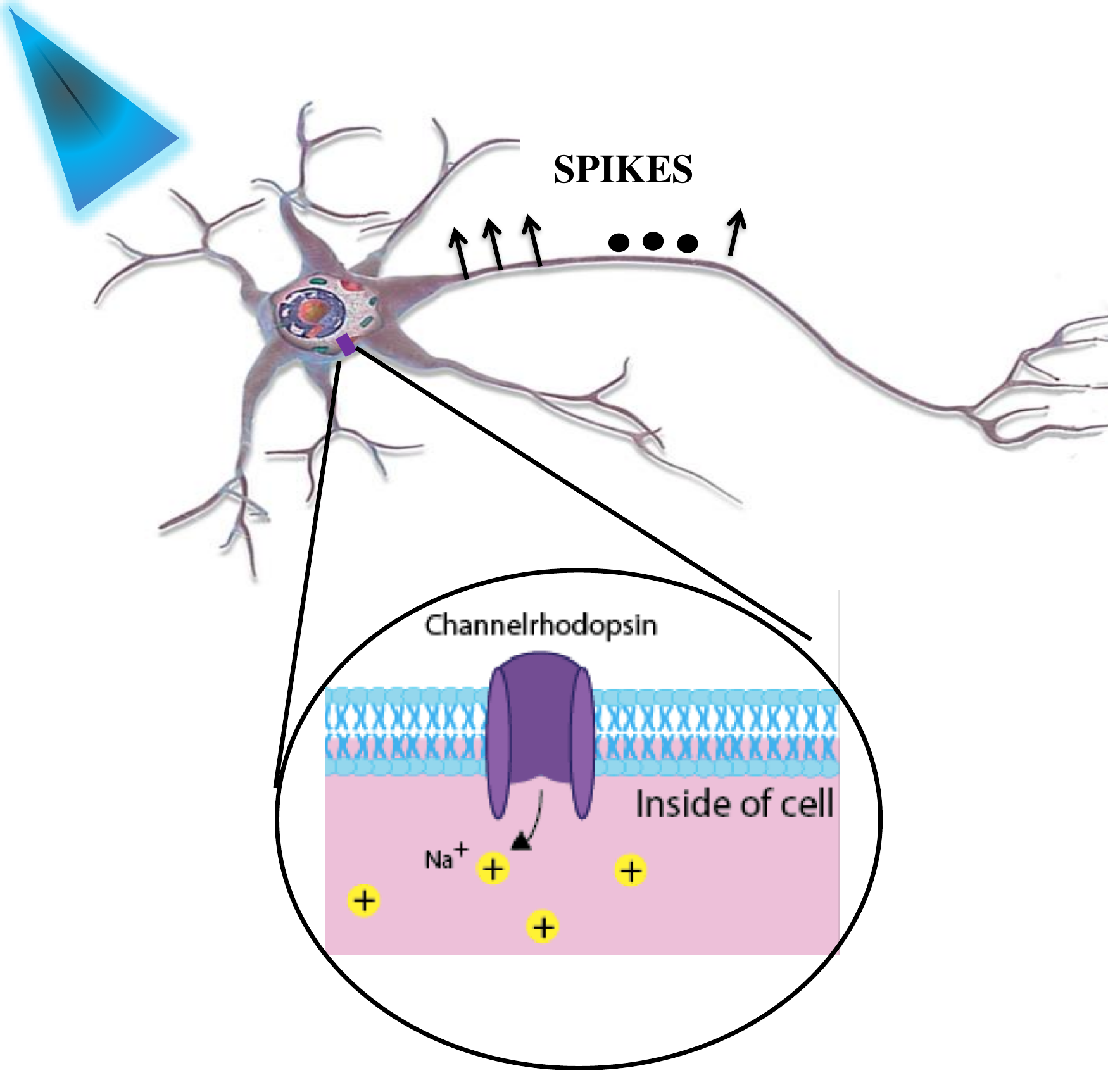}}
	
	\caption{Optogenetic neuronal inhibition and excitation using light responsive proteins called opsin \cite{deisseroth2011optogenetics}.}
	\label{fig:optogen}
\end{figure*}

The sensory feedback from the body, which is processed by neurons in the spinal cord, is crucial in controlling the motor commands \cite{courtine2009transformation}. Hence, the idea of stimulating the spinal cord just enough to make it sensitive and responsive to the sensory input is studied in \cite{harkema2011effect}. An array of $16$ electrodes is implanted in epidural space of the spinal cord of an individual with a clinically complete motor injury. These electrodes are placed over the identified extensor and flexor hotspots, and a pulse generator is used to control the stimulation parameters. Different set of stimulating parameters are then sent to the pulse generator to observe the response of the patients body to the stimulation. The epidural stimulation in \cite{harkema2011effect} resulted in locomotor-like patterns and full weight-bearing standing with assistance provided only for balance. Instead of selection of stimulating parameters by a physician, which is done in \cite{harkema2011effect}, the recorded signals from brain or spinal cord can be processed by a processing unit to set these parameters. Using brain signals to control the epidural stimulation is studied in \cite{wenger2014closed,capogrosso2016brain}, where locomotor movements, namely walking and climbing, are achieved in paralyzed monkey and rat, respectively. 

In \cite{zimmermann2014closed}, intra-spinal micro-stimulation is used to control hand movement in monkeys whose hands were temporarily paralyzed. This stimulation method provides more natural recruitment order of motor units and reduces the number of required electrodes and controllers compared to epidural stimulation. However, higher risk of tissue damage exists due to implantation of electrode into the spinal tissue. Implanted electrodes retain a scar tissue, and long-term stability and safety of the implanted electrodes are some of the major challenges~\cite{bamford2010effects}.

By using current electrical stimulation techniques in either epidural stimulation or intra-spinal micro-stimulation, cell-specific activation of spinal cord's motor circuitry is challenging due to interference from non-target neuron populations. In this respect, optogenetics can be a key to control the limbs by activating or suppressing the specific population of neurons in the spinal cord as shown in Fig.~\ref{fig:optogen} \cite{caggiano2014rostro,caggiano2016optogenetic}. Open issues include finding encoding methods for motor commands recorded from the brain in order to deliver meaningful signals to the spinal cord and developing proper light delivery techniques for spinal cord \cite{montgomery2016beyond}. The former requires identifying contributions of different neuron populations on the motor control. Regarding the latter, current research focuses on the RF-powered wireless systems \cite{park2015soft,montgomery2015wirelessly} since employing optical fibers in the spinal cord is not possible. This is contrary to the case of the brain since penetrating an optical fiber in the spinal cord requires the severing of white matter tracts, which carry high information density and have minimal redundancy; thus, local damage can have global consequences \cite{montgomery2016beyond}. Another approach could be designing an ultrasound-powered optogenetic stimulator system, such as those in \cite{weber2016miniaturized, arbabian2016sound}, which can reach deeper layers in the spinal cord, to employ them in the spinal~cord.

\subsection{Artificial Neurons}

\begin{figure*}[t]
	\centering
	\includegraphics[width=16cm]{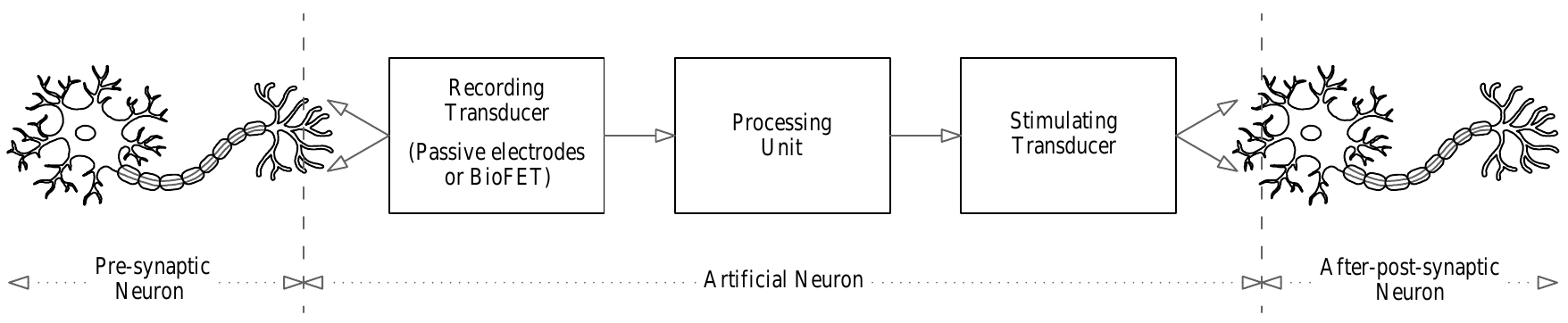}
	\caption{The general architecture of an AN, which fulfils processing tasks similar to its biological counterparts, completing the nervous pathways.}
	\label{fig:AN_Gen}
\end{figure*}

Another potential direction for the treatment of SCI could be the use of artificially-created neurons to replace the injured biological neurons. If perfectly implemented, such an approach can potentially reverse the loss of any function occurred due to SCI; however, the associated challenges are not negligible by any means either. The general architecture of an artificial neuron (AN) is shown in Fig.~\ref{fig:AN_Gen}, where a pre-synaptic neuron connects to a processing unit by means of a transducer. The AN performs processing actions similar to its biological counterparts, and thus completes the nervous pathways.

Some key characteristics required by a functioning AN that may replace biological neurons are outlined below:

\begin{itemize}[leftmargin=*]
	\item \textit{Plasticity:} The AN should be able to change synaptic weights in a plastic manner since plasticity forms the basis of memory and cognitive functions of the nervous system and is observed in all neurons in a nervous system. It equips a network of ANs with the capability of mimicking the observed behavior in networks of biological neurons of changing activity patterns according to durable change in input stimuli~\cite{pfingst1977response}.
	
	\item \textit{Implementation scale:} The scale of implementation decides the level on which the system behaves similar to a biological neuron. From the perspective of a single neuron, the scale can either be synaptic or neuronal. On the synaptic scale, individual synapses can be identified and interfaced to synapses from other neurons, whereas on the neuronal scale, the neuron as a whole may be interfaced. Both systems require a different set of interfacing techniques but the complexity and control available in AN systems will increase against a decrease in the scale.
	
	\item \textit{Biocompatibility:} Same as NIS, the AN needs to provide lifelong stability, which requires biocompatibility and cytocompatibility of the device, i.e., the device must integrate properly within the body and neural tissues, and provide the prolonged maintenance of the desired performance \cite{larsson2013organic}.
	
	\item \textit{Physical dimensions:} The size of ANs need to be defined based on the biological neurons that they will be integrated with. Thus, their dimensions must have parts on the scale of micrometers to be compatible with thicknesses of axons and dendrites of biological neurons, which need to be integrated together with other parts of an AN of larger dimensions.
	
	\item \textit{Energy requirements:} Energy self-reliance is another major issue for any implanted system. Use of batteries may not be a viable solution for long-term operation. The energy required by an AN can be harvested from the body heat or blood pressure by creating gradients within the body \cite{beeby2006energy}; however, other novel methods can be proposed. For instance, ANs may be designed such that they can harvest energy directly from the energy sources of the body, such as glucose, and their efficiency should also be considered.
	
	\item \textit{Amplification:} Processing of biological inputs may require the use of amplifications since the transmission needs to be carried over some distance and the typical energy of signals is on the order of $\mu$J. Such amplifiers must have low noise, low power, and high precision characteristics.    
	
	\item \textit{Implanting process:} This process is almost as critical as the development of ANs. Considering the scale of neurons, operation by a human surgeon is not a viable option. The implant technologies may require the use of robotic surgeries that are capable of operating at micrometer scales, which is already a realized technology by Neuralink~\cite{pisarchik2019novel} for the case of neural electrode implantation. In such cases, the system should be intelligent since the sheer number of decisions will be too large for a human to process. The system would need to automatically identify different sections of biological neuron by means of imaging, and proceed with the surgery by implanting the appropriate AN sections accordingly. Thus, apart from the robotic arms and surgery tools, the implantation requires significant ML and artificial intelligence (AI) systems.
	
	The implantation problems elaborated above can be mitigated by using systems that deploy themselves by means of self-organization.
	
\end{itemize}

The characteristics/requirements of ANs pose significant challenges; however, once developed, the benefits of such systems are also unlimited. Apart from the treatment of SCI, they may be utilized to develop computing systems similar to the human brain. Additionally, they may be used as bio-cyber interfaces for applications targeting the nervous system.

To date, there is no working prototype of an AN that may be considered as a complete replacement for a biological neuron. Thus, we discuss some major studies that are approaching to the problem in a similar manner. Two major directions reported, namely neuromorphic and biomimetic neurons, are explained below: 
\subsubsection{Neuromorphic neurons}
Neuromorphic neurons are based on electronic circuits, 
which particularly perform neuromorphic computation, where very-large-scale integration (VLSI) of individual neurons or synapses are utilized in a similar manner to the human brain \cite{mead1990neuromorphic}.

The biggest advantage of neuromorphic systems is that they include plasticity as the basis of their models; however, implementation of ANs by neuromorphic methods faces several bottlenecks, such as huge size, high-power requirements, and interfacing issues. Even for some of the newer devices that use low-power and have small sizes, interfacing issues remain since neuromorphic systems are not designed with neural interfaces \cite{saighi2015plasticity}. 
\subsubsection{Biomimetic neurons}
Biomimetic circuits mimic biochemical processes using synthetic materials \cite{otero2012biomimetic}. Since most communication between biological entities occurs by means of molecule exchange and biochemical reactions of these molecules, biomimetic circuits are ideal candidates for direct interfacing with biological systems. They are employed to realize applications, such as artificial muscles, smart membranes, biological transducers, and brain-machine interfaces \cite{otero2012biomimetic}.

The basic function of a neuron is realized by means of enzyme-based amperometric biosensors and organic ion pumps in a biomimetic fashion by \cite{simon2015organic}. Although this work is bio-compatible and the detection of bio-molecules is achieved on the scales similar to those of biological neurons, many of the characteristics expected from an AN are not met. For instance, there is no plasticity in the system, macro-scale dimensions are used \cite{simon2015organic}, and the system is tested in vitro without any interface to the biological neurons.

Similar studies that include plasticity and direct interface with biological neurons may be the first step towards the development of organic biomimetic ANs. Size reduction and in vivo testing may follow afterwards. Another direction could be the use of plasticity designs from neuromorphic systems based on biomimetic implementations. This may be the approach that is most successful and has the least time-to-market. 

\section{Proposed Approaches of the ICT-based Treatment of SCI}
In this section, we introduce two potential approaches for the treatment of SCI by considering the current literature and the desired characteristics of NIS and ANs.
\subsection{Neural Interface Systems with Enhanced Feedback}
The main technical challenges in using an NIS to enable the ICT-based treatment of SCI are over-viewed below:
\begin{itemize}[leftmargin=*]
	\item \textit{Providing somatosensory feedback}: It is shown that visual monitoring of a limb's motion can partially correct the shortcomings in movement of patients with large-fiber sensory neuropathy. These subjects face slow and uncoordinated movement as a result of the somatosensory feedback for normal motor control \cite{ghez1990roles}. Hence, patient's vision provides a feedback to the NIS. However, the effect of visual feedback on movements is not as fast as proprioceptive system, which then causes instability in movement \cite{hatsopoulos2009science}. Thus, utilizing proprioceptive feedback or a feedback from spatio-temporal characteristics of limb's in an NIS can assist patients to perform more natural movements and also feel stimulated limbs or neuroprosthesis \cite{wenger2014closed,ramos2012proprioceptive}. As an example, stimulating peripheral afferent nerves in the limb's stump with intrafascicular electrodes \cite{dhillon2005direct,horch2011object} or cuff-like electrodes \cite{tyler2002functionally} is studied in the literature for providing the natural sensation of missing limbs in amputee subjects. However, these methods are not applicable in paralyzed patients, who have also lost the sensory pathway. In such cases, the tactile and proprioceptive feedback must be provided by artificially stimulating the sensory cortex. 
	\item \textit{Appropriate neural recording technique, spinal cord, and brain stimulation}: 
	Lifelong stable neural signal recordings and stimulation are required for which the important factors are reliable chronic recordings, biocompatibility and cytocompatibility of the device, the ability of the device to integrate properly within neural tissues, and the prolonged maintenance of desired electrical properties. Furthermore, both recording and stimulating devices must have high spatio-temporal resolution to handle the required degree of freedom for restoring function after the SCI. Graphene is a bio-compatible 2D material, and its sensitivity to different ions can be modified by various etching methods and functionalization \cite{rollings2016ion}. Hence, it can be used for sensing fluctuations in specific ion concentrations. Graphene is electrically conductive, and it can be patterned into nanoscale, electrically disconnected, conductive patches via hydrogenation, etching or fluorination \cite{elias2009control,nair2010fluorographene}. This makes it suitable for implementing high resolution neural interfaces which can both generate and differentiate the highly localized fluctuations in ion concentrations.
	\item \textit{Fully implantable system}: To offer more comfort for patients and reduce the risk of inflammation, the NIS must be fully implantable. Moreover, the system must provide the required bandwidth for transmitting data between its different components while considering the power consumption.
\end{itemize}

\begin{figure*}[t]
	\centering
	\includegraphics[width=18cm]{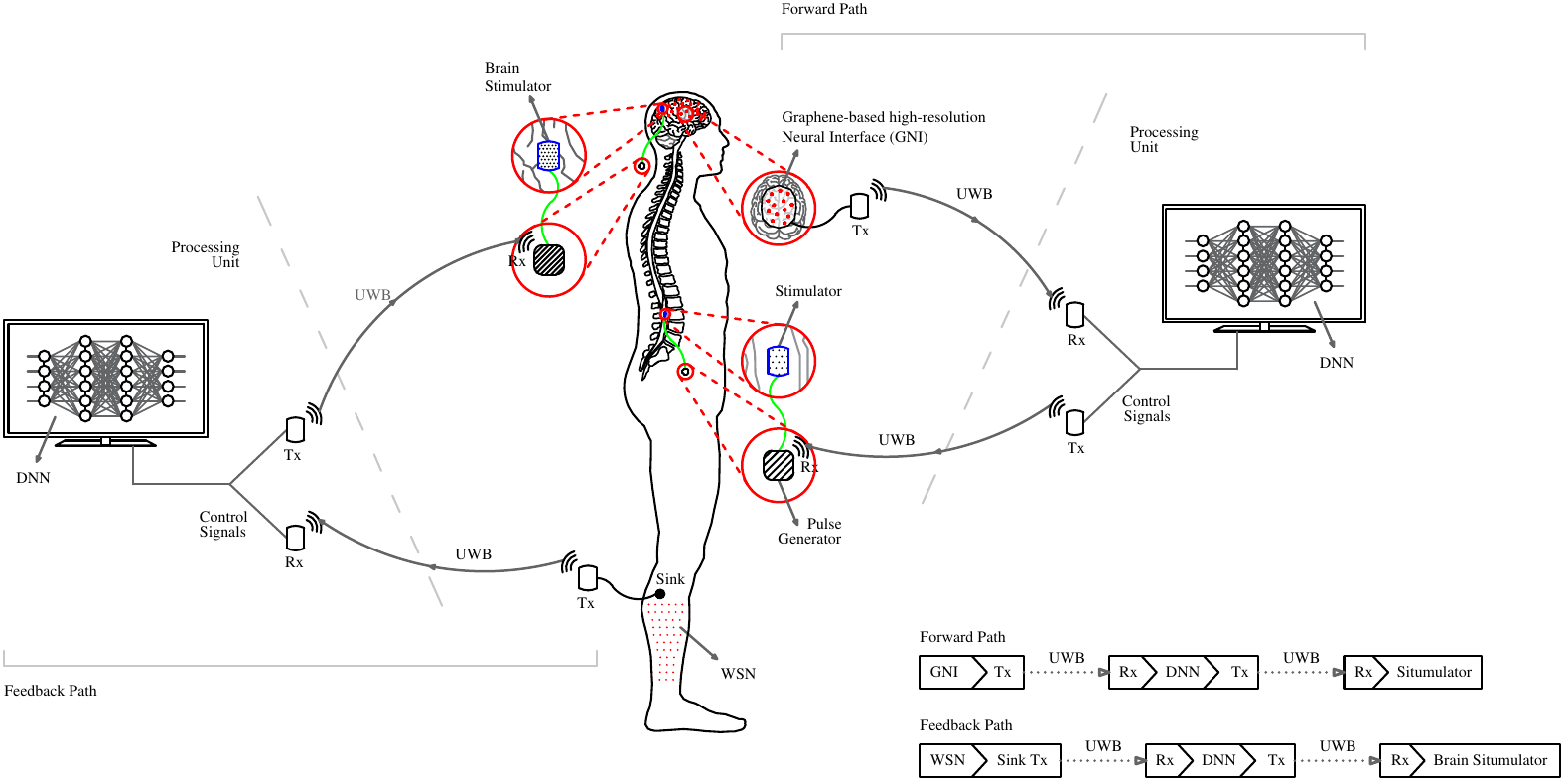}
	\caption{The neural interface system with enhanced feedback, EF-NIS, that utilizes brain-recorded neural signals and stimulates the spinal cord to restore the motor functions of the paralyzed patients.}
	\label{NIS}
\end{figure*}

To address the above-mentioned challenges, we propose a comprehensive framework for realizing an NIS with enhanced feedback, \textit{EF-NIS}, that utilizes the neural signals recorded from the brain and stimulates the spinal cord for restoring motor functions in the paralyzed patients as shown in Fig.~\ref{NIS}. Components of the proposed EF-NIS are explained below:

\subsubsection{Recording device and the communication interface}
Graphene-based devices are particularly attractive for neural interfacing as graphene promotes neurite growth \cite{li2011promotion}, do not alter the behavior of target neurons \cite{fabbro2016graphene}, is flexible and transparent, which allows neuroimaging/optogenetic stimulation concurrent to neural electrophysiology \cite{kuzum2014transparent}. Moreover, graphene is highly conductive and electrically sensitive, ideal for lowering SINR \cite{blaschke2017mapping}. To this aim two set of devices are studied in the literature. Graphene based Micro-Electrode Arrays (MEAs) provide very dense arrays with submicron electrodes \cite{lu2018graphene, lu2016flexible} while having easy and reliable fabrication process. However, MEAs do not provide any on-site amplification thus implying noise in links severely effects SINR and the recorded signal needs complicated post processing. On the other hand, graphene Field Effect Transistors (GFETs) provide on-site amplification of the recoded signal, thus, high SINR values \cite{hess2013graphene}. In addition, biochemical sensing of neurotransmitters, e.g., glutamates, is possible by functionalization of the surface \cite{huang2010nanoelectronic}. By selecting the appropriate receptor molecules, we can mimic realistic postsynaptic membrane potential decay times, which leads to implementation of both temporal and spatial summation of postsynaptic membrane potential. Thus, the graphene-based high resolution neural interface (GNI) is suggested to record the brain activities of the patient from primary motor cortex. \color{black} This recording device is implanted together with a processing unit and a transmitter, which reports the recorded signals in a wireless manner. 
 
The bandwidth demand of the wireless transmission channel between the brain implant and the external processing unit is correlated with the number of recording channels employed. For better estimation of brain commands, we need densely distributed recording sites that require high data rates (on the order of Mbps) \cite{zhang_2016}. On the other hand, the transmitter should dissipate low energy because even 1$^{\circ}$C of temperature increase can damage neural tissues \cite{kim2007thermal}. Considering these limitations, ultra-wide band (UWB: $3.1$-$10.6$ GHz) systems, which can provide high data rate within allowable power dissipation range with the minimal chip area \cite{6572893}, is the most suitable solution \cite{5201317}. Therefore, the communication interface between the brain implant-processing unit of the EF-NIS consists of a simple UWB transmitter at the brain implant and a more complex UWB receiver at the processing unit \cite{6594741}. Furthermore, compression techniques with efficient implementation, similar to one in \cite{7469363}, can be utilized to reduce data rate without considerably increasing the complexity.
\color{black}

\subsubsection{Processing unit and the spinal stimulator interface} The processing unit performs two tasks: (i) decoding the intent of the subject from neural signals; (ii) generating control signals for spinal cord stimulation. This unit will be placed in a rechargeable and mobile device. Hence, it needs a wireless transmitter to transfer the stimulation parameters to the site of spinal cord stimulation. The transmission between the processing unit and the spinal cord neural stimulator can be performed at lower data rates compared to the brain implant-to-processing unit link since the dimensionality of the transmitted signal reduces. Additionally, the receiver at the neural stimulator must operate with low energy \cite{kim_2006}. Area- and power-efficient UWB transmission is suitable for this link \cite{ture2018area,tokgoz2018120gb}. Moreover, it is also advantageous in terms of interference \cite{7420751}. 

The received control signals at the spinal cord stimulator are then used for setting the stimulation parameters. The GNI is able to both recognize and generate fluctuations in concentration of specific molecules in a local area. Hence, it will also be used for high resolution stimulating of nerve cells located in the spinal cord.

\begin{figure}[h]
	\centering
	\subfigure[Graphen electronic tattoo sensors \cite{kabiri2017graphene}.]{\label{fig:Gtattoo}\includegraphics[width=83mm]{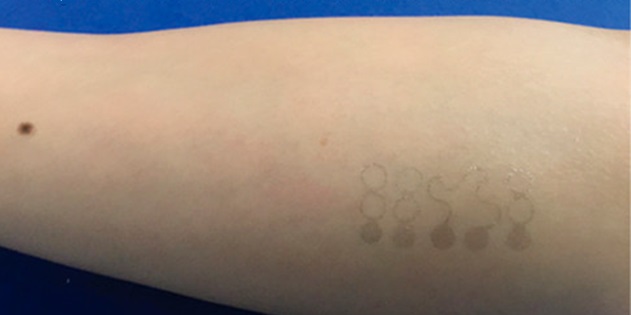}}
	\vspace{0.7mm}
	\subfigure[A transparent graphene-based piezoelectric motion sensor \cite{lim2015transparent}.]{\label{fig:Graphene}\includegraphics[width=83mm]{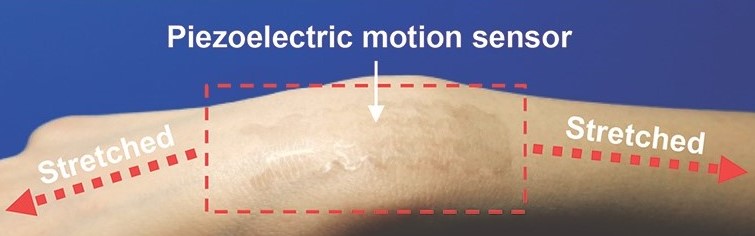}}
	\vspace{0.7mm}
	\subfigure[Flexible motion sensor using gold thin films \cite{lim2015transparent}.]{\label{fig:Gold}\includegraphics[width=83mm]{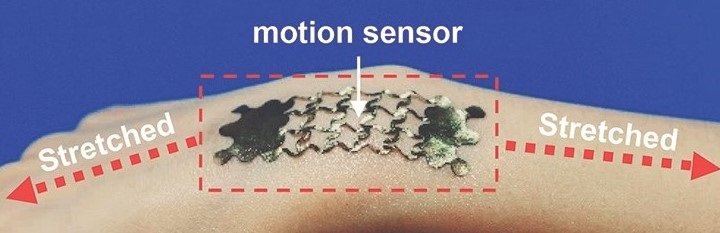}}	
	\caption{Examples of flexible tattoo-based and skin-like sensors.}
	\label{fig:Tattoo_Skin_Like}
\end{figure}

\subsubsection{Muscle to brain communication interface (Feedback)}
In literature, there is a variety of wearable sensors for monitoring the body activity. Among those, especially flexible tattoo-based and skin-like sensors are profitable due to their soft and ultra-thin designs that provide comfort for users. Furthermore, they are generally made of bicompatible materials, such as graphene \cite{kabiri2017graphene,lim2015transparent, kang2017graphene,zhang2016human} and gold \cite{kao2016duoskin}, as shown in Fig.~\ref{fig:Tattoo_Skin_Like}, which makes them suitable for long-term use. These sensors are also capable of recording multiple sensory modalities, such as muscle activity and touch \cite{kabiri2017graphene}. For the communication interface, wired, wireless or hybrid approaches can be followed. For high density of sensors, which is the case of multiple modalities being recorded, we can use wired sensor arrays, whose recordings are gathered at the sink node, which has a direct wireless communication interface with the processing unit or with the brain implant. Otherwise, communication modules of the sensors can be integrated to Near Field Communication (NFC) and Radio Frequency Identification (RFID) technologies, which are enabled by flexible thin antennas, such as graphene-based antenna shown in \cite{huang2015binder}.

An alternative to wearable sensors is ultrasound-based implantable neural dust, which is validated in rat peripheral nerves and is shown to be capable of recording neural activity in the peripheral nervous system \cite{seo2016wireless}. This system uses backscattering and other advantageous aspects of ultrasound, such as low attenuation through the tissues and low energy consumption compared to RF technologies currently employed in implantable devices \cite{santagati2017experimental}. In addition, further miniaturization is needed in this system similar to the aforementioned flexible sensor-based systems. 

The collected sensory information is then transmitted to the processing unit to generate the parameters for somatosensory cortex stimulation. For this aim, the responsible regions of somatosensory cortex for sensory feedback from each leg should be identified. This task can be achieved by collecting tactile and muscle activities in healthy subjects, and recording the corresponding activities in somatosensory region during walking. Then, a DNN is trained to estimate the activity of the somatosensory cortex according to the collected sensory information. Since we consider the SCI injuries, in which the sensory pathway is also damaged, EF-NIS utilizes this information to generate stimulation parameters that activate similar somatosensory regions in patients. 

In the next step, the processing unit transmits the stimulation parameters to the controller unit that is surgically placed over the pia mater, i.e., the delicate innermost layer of the meninges that is the membranes surrounding the brain and the spinal cord. The transmission is again performed through the UWB communication interface. The controller sends the control signals to the graphene-based high-resolution neural interface, located over the somatosensory region, to stimulate the brain through manipulating ionic concentration levels. A possible approach to achieve the manipulation of ionic concentration levels, graphene membranes with tunable sub-nanometer pores are a suitable candidate, which are shown to help to facilitate selective ion transport through them~\cite{o2014selective}.

\subsection{Self-Organizing Artificial Neurons}

Although the preceding neural interface is relatively easy to implement and may return a high degree of motion to the patients, the movements will not be precise. In similar, although the nervous system's information processing capacity increases by the use of an NIS, its communication rate is not as high as before the injury. Before an SCI, being normally wired, the nervous system can be assumed to operate at optimum connectivity with maximum communication rate. However, after the SCI, while using an NIS, neural information can only be communicated at limited neural regions, a setup far from the optimum connectivity, which results in diminished communication rate. This materializes itself, for instance, as limited stimulation of some estimated areas of the spinal cord for the motor movement. Apart from limitations in establishing connectivity, operating through interfaces located far away, e.g., the cerebral cortex, from the spinal injury site, where the lost connectivity to be reestablished resides, additionally suffers from the delocalization of neural signals as they propagate. Thus, establishing a simple lost connection at the injury site may correspond to collecting readings from the multiple sites of cortex together with some signal post-processing. However, the delocalization process, coupled with a lack of complete neural signals, results in unrecoverable information loss, and the lost connection will not be recovered in full effect. On the other hand, delocalization of neural signals is not an issue if one opts to work at the injury site.

A solution to this problem could be the retrieval of connections lost at the injury site via the self-organizing ANs. The connectivity, and consequently communication rates, could be (at least theoretically) brought back to the pre-injury levels by employing artificial nervous connections across the injury site, where the signals are already localized in specific positions. An architecture, where ANs with neuromorphic properties are suspended in a hydrogel, could be employed for such purpose. The hydrogel-suspended network of ANs can form bridges across the injury site and make new nervous connections with biological and artificial neurons via self-organization. A representative diagram is shown in Fig.~\ref{fig:AN_Block}. Below, we discuss the major properties of the proposed self-organizing AN:

\subsubsection{Plasticity}
In order for an AN to adapt to its environment, plasticity should be introduced in the system and realized by utilizing neuromorphic designs. The firing rate of each AN depends on the signals it receives from synapses with other neurons that may be both biological or artificial. Each synapse of the AN is associated to a variable $Weight$ that models plasticity of the synapse and retains the weight of a particular synapse \cite{gerstner2002spiking}. The $Weight$ may function to strengthen or weaken depending on the synaptic connection type (excitatory or inhibitory). The overall effect of plasticity is the introduction of memory in the system, which is a key characteristic of the nervous system.
\subsubsection{Self-organization}
\begin{figure*}[t]
	\centering
	\subfigure[Before self-organization.]{\label{fig:a}\includegraphics[width=50mm]{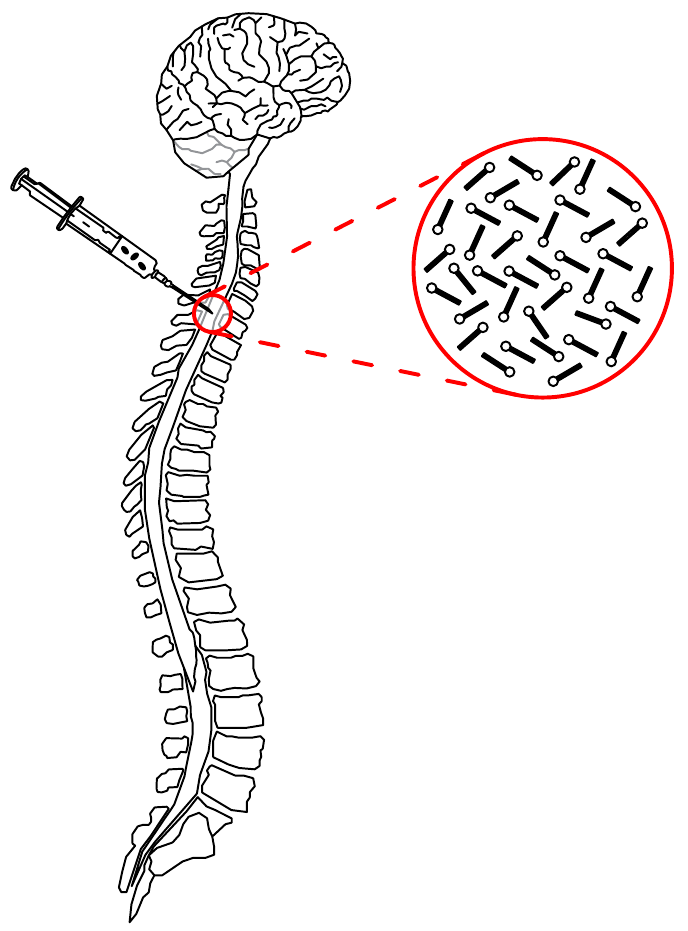}}
	\subfigure[After self-organization.]{\label{fig:b}\includegraphics[width=50mm]{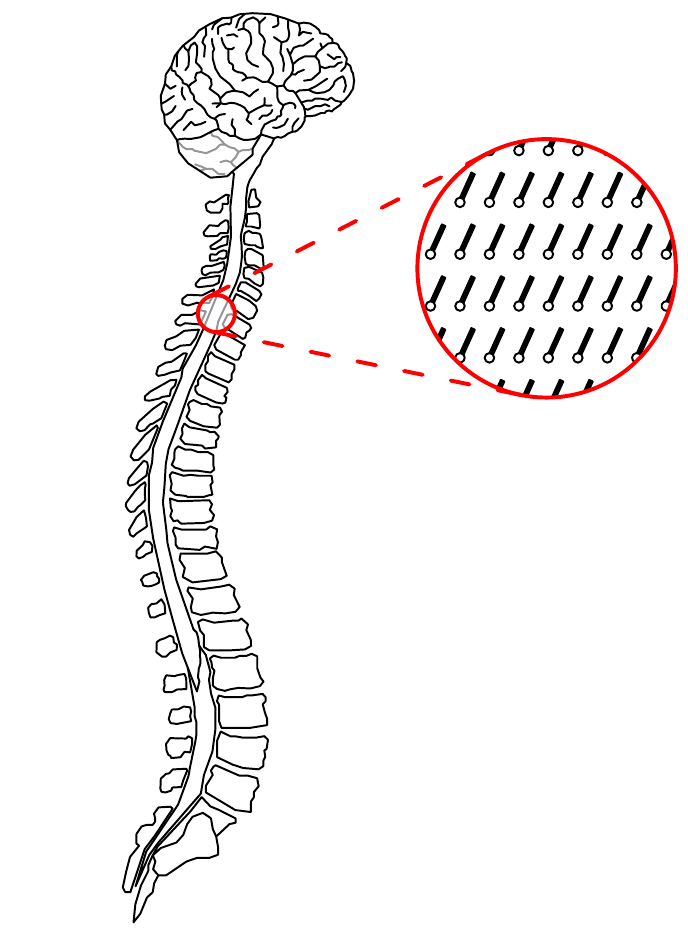}}
	\caption{Diagram of the self-organizing ANs suspended in a hydrogel, where AN bridges form across the injury site over time.}
	\label{fig:AN_Block}
\end{figure*}

Since functional ANs are required to interface with biological neurons, their dimensions are dictated by the task. If ANs are communicating on axonal and somatic levels, their sizes will be micro-scale whereas if ANs aim to communicate on synaptic scales, their sizes should be nano-scale. As discussed earlier, the implantation of AN systems propose a major challenge since surgery on the micro/nano-scale itself is an open problem. This can be bypassed by considering systems that use self-organization to create the AN bridge across the nervous injury. Such a system is shown in Fig.~\ref{fig:AN_Block}, where AN bridges form across the injury site over~time.

Self-organization is a hot research area, where local interactions between individual entities give rise to the overall structure of the system. The simplest approach towards this goal is the use of electric or magnetic fields. Since the action potentials of the nervous system are themselves electrical signals, self-organization can be based on its use as the attractor field. The concept of alignment of microstructures by means of anisotropic electric field potentials has already been demonstrated for the case of metallic microcoils suspended in PDMS~\cite{li2017remarkable}. Graphene nanosolenoids discussed in \cite{xu2015riemann} may provide the basis for an implementation in this regard, both in terms of the required size and low energy requirements and magnetic properties. The nanosolenoids, however, are not sufficient by themselves for self-organization, because the nervous signals are not continuous in nature and the AN bridge may fall apart if a signal is absent for some time. To ensure long-term endurance of the structures, we either need mechanical structures, such as hinges \cite{tian2010design}, or the use of magnetic fields external to the body \cite{kralj2015magnetic}.
\subsubsection{Energy Harvesting}
For the system to be completely independent, ANs need to extract energy from their vicinity. This energy can be scavenged within the body \cite{beeby2006energy} or delivered through external sources using wireless power transfer (WPT) \cite{khan2020nanosensor}. 

Energy harvesting (EH) within the body can either be based on biochemical or biomechanical processes \cite{akhtar2017energy}. Chemical reactions within the body, such as glucose uptake and food digestion, can be considered as the sources for biochemical EH mechanisms. Here, glucose and oxygen come to the forefront and are used extensively compared to several other biochemical pairs due to their large quantity and ubiquity in human body \cite{yang2020challenges}. When the glucose is burnt with oxygen in an implantable enzymatic biofuel cell, the free electrons are released, which generates the electrical energy. Typical to any chemical reaction, the rate of this oxidation can be increased by such substances called catalysts, the characteristics of which affect the level of power generation. The maximum power density that can be gathered by glucose oxidation is $180$ $\mu$W/cm$^2$ \cite{akhtar2017energy}.

On the other hand, biomechanical EH can either be based on voluntary body functions, such as movement, or involuntary functions, such as blood flow, breathing, and heartbeat, which create utilizable vibration. The sources of locomotion, i.e., any voluntary movement, can be converted into electricity using piezoelectric and electrostatic materials or nanogenerators. Piezoelectrics can be also used for utilizing the involuntary body functions in addition to nanowires and electromagnetic induction. It is possible to create up to $6$ $\mu$J/cm$^3$ of power density out of blood pressure, which is $1.2$ $\mu$W/cm$^2$ for heartbeat. Depending on the location of sources, hybrid EH mechanisms can be also implemented to increase the likelihood of continuous energy reception, allowing a source to compensate for another in case of failure \cite{akan2017internet}. The key advantage of EH within body is the fact that the designed system can be independent of external influences, allowing subjects to move more freely.

WPT can be realized by providing power to the system externally by means or either inductive or magnetic coupling, or by means of electromagnetic radiation \cite{wagih2020real}. Apart from reducing the challenges of designing EH mechanisms in an already complex scenario of SCI treatment, if an external source is selected, the same may be used for the self-organization of the AN devices, and as a readout interface of the AN network. Finally, it should be noted that EH is a very active area of research. New efforts are focused on reducing the power requirements of networks by using energy-efficient protocols, conservation schemes, and novel topologies. Future work can further elaborate on the directions overviewed here.

\section{Further applications of the ICT-based treatment of SCI}

By revisiting the current treatment techniques of SCI from an ICT perspective, we can provide rigorous optimization frameworks for the improvement of these techniques. However, apart from SCI treatment, such ICT-based techniques can have applications in a number of other scenarios. Thus, in this section, we discuss some key applications that are realizable as byproducts of ICT-based SCI treatment techniques.
\subsection{The Internet of Bio-Nano Things} The Internet of Bio-Nano Things (IoBNT) is an emerging subfield of IoT, which focuses on developing artificial biological/bio-inspired nano-scale devices capable of communicating with each other and their surroundings, e.g., biological entities/parts or trace biomolecules from biological entities/parts, to establish predefined tasks at the nano-scale in a biological environment, with envisioned applications ranging from intra-body sensing~\cite{kuscu2018maximum} and actuation networks to environmental control of chemical and biological agents and pollution \cite{akyildiz2015internet}. In the case of intra-body applications, experience gained from techniques developed for SCI treatments, in particular, molecular NIS, can provide the means for establishing communication with nano-networks deployed into the body. Such applications include remote monitoring of the body for various vectors and automatized targeted drug delivery upon detection of anomalies in observed data. 

\subsection{Neurophysiology/Physiology} Large amounts of data can be retrieved from the nervous system by using the interfaces of the proposed SCI treatments. These data can be processed to obtain a mapping between the architecture of the underlying network and its function, and contribute vastly to our understanding of biological neural networks. On the other hand, NIS-like systems equipped with molecular sensors capable of sensing specific biomolecules can be used to interface other biological systems, such as cardiovascular and lymphatic systems. Data collected from such an endeavor will significantly contribute to our understanding of human physiology. For instance, we can study nervous-related diseases, such as Alzheimer's disease, multiple sclerosis, and Parkinson's disease, in unprecedented detail using electrical/molecular NIS as interfaces to the central nervous system. This may lead to novel disease diagnosis and treatment techniques as well as the identification of key factors that cause such conditions.
\subsection{Neuromorphic Computing} A firm understanding of the relation between the architecture and function of a neural network would allow us to design our own artificial neural networks (ANNs, as a subtopic of ML and AI) with specialized functions, and employ these architectures for neuromorphic computation at large scale~\cite{furber2016large}. Neuromorphic architectures process data in a highly parallel manner as opposed to the serial structure used by von Neumann architectures employed by contemporary computers. This allows ANNs to process data at comparatively very low frequencies, and therefore to be highly energy efficient. 

One application of functional ANNs lies in their integration into neural interfaces for on-body preprocessing of collected data, which is necessary as the volume of raw data collected by high spatio-temporal resolution interfaces will be so large that transmission of this data without any preprocessing is doomed to have heating problems, which is not desired for an on-body device. Thus, we will have a positive feedback loop: Understanding gained from neural interfaces will advance our capability of neuromorphic computing, which, in return, will pave the way for more advanced interfaces.
\subsection{Augmented Humans (Cyborgs)} The use of external machinery to enhance capabilities of a human body is not new. In fact, prosthetics are widely used to help people with missing limbs to regain some functionalities lost due to suffered injury. With advancements in SCI treatment technology, it is expected that such prostheses will be upgraded from being mere mechanical tools to interactive electronic devices capable of communicating with human nervous system as the so-called neural prostheses~\cite{cognolato2020gaze}. Interesting applications in such directions may include extrasensory feedbacks to achieve new sensations (e.g., augmented reality~\cite{garry2019cyborgs}), mechanical body suits that allow superhuman strength, direct integration of gaming to the human body, among others.

\section{Conclusions}
SCI are serious health conditions that adversely affect the life of a plethora of patients around the world. Existing biological methods not only aim to re-establish axonal connection by promoting nerve regeneration but also minimize harmful toxicities of the injury mechanisms. However, to date, no existing treatment technique has shown significant success. Among the ICT-based systems, NIS simply bypass the injury site by decoding the brain signals and providing them to the recipient muscles. On the other hand, employment of ANs aims to compensate the functional loss by replacing the injured neurons with engineered neurons. Considering these, we propose two novel approaches, namely EF-NIS (NIS with enhanced feedback) and self-organizing ANs, and identify key areas that may be most important in this regard. For NIS, apart from the need of high resolution recording and stimulation setups, we identify that the feedback from impaired limbs back to the brain is also very essential. For ANs, we propose the use of self-organization to tackle the transplant issues in a smarter way and also EH to make the devices fully independent. The development of these treatments will have a great impact on a number of application areas, such as IoBNT. 
\bibliographystyle{IEEEtran}
\bibliography{References}

\begin{thebibliography}{100}
\providecommand{\url}[1]{#1}
\csname url@samestyle\endcsname
\providecommand{\newblock}{\relax}
\providecommand{\bibinfo}[2]{#2}
\providecommand{\BIBentrySTDinterwordspacing}{\spaceskip=0pt\relax}
\providecommand{\BIBentryALTinterwordstretchfactor}{4}
\providecommand{\BIBentryALTinterwordspacing}{\spaceskip=\fontdimen2\font plus
\BIBentryALTinterwordstretchfactor\fontdimen3\font minus
  \fontdimen4\font\relax}
\providecommand{\BIBforeignlanguage}[2]{{%
\expandafter\ifx\csname l@#1\endcsname\relax
\typeout{** WARNING: IEEEtran.bst: No hyphenation pattern has been}%
\typeout{** loaded for the language `#1'. Using the pattern for}%
\typeout{** the default language instead.}%
\else
\language=\csname l@#1\endcsname
\fi
#2}}
\providecommand{\BIBdecl}{\relax}
\BIBdecl

\bibitem{akan2016Fundamentals}
O.~B. Akan \emph{et~al.}, ``Fundamentals of molecular information and
  communication science,'' \emph{Proceedings of the IEEE}, 2016.

\bibitem{Waxman_2013}
S.~G. Waxman, \emph{Clinical Neuroanatomy}.\hskip 1em plus 0.5em minus
  0.4em\relax New York: McGraw-Hill, 2013.

\bibitem{james2019global}
S.~L. James \emph{et~al.}, ``Global, regional, and national burden of traumatic
  brain injury and spinal cord injury, 1990--2016: a systematic analysis for
  the global burden of disease study 2016,'' \emph{The Lancet Neurology},
  vol.~18, no.~1, pp. 56--87, 2019.

\bibitem{SC_facts}
\BIBentryALTinterwordspacing
``Spinal cord injury (sci) 2016 facts and figures at a glance,'' \emph{The
  Journal of Spinal Cord Medicine}, vol.~39, no.~4, pp. 493--494, 2016, pMID:
  27471859. [Online]. Available:
  \url{http://dx.doi.org/10.1080/10790268.2016.1210925}
\BIBentrySTDinterwordspacing

\bibitem{fetz1999real}
E.~E. Fetz, ``Real-time control of a robotic arm by neuronal ensembles,''
  \emph{Nature neuroscience}, vol.~2, pp. 583--584, 1999.

\bibitem{wessberg2000real}
J.~Wessberg \emph{et~al.}, ``Real-time prediction of hand trajectory by
  ensembles of cortical neurons in primates,'' \emph{Nature}, vol. 408, no.
  6810, pp. 361--365, 2000.

\bibitem{serruya2002brain}
M.~D. Serruya \emph{et~al.}, ``Brain-machine interface: Instant neural control
  of a movement signal,'' \emph{Nature}, vol. 416, no. 6877, pp. 141--142,
  2002.

\bibitem{taylor2002direct}
D.~M. Taylor \emph{et~al.}, ``Direct cortical control of 3d neuroprosthetic
  devices,'' \emph{Science}, vol. 296, no. 5574, pp. 1829--1832, 2002.

\bibitem{guo2014encoding}
Y.~Guo \emph{et~al.}, ``Encoding of forelimb forces by corticospinal tract
  activity in the rat,'' \emph{Frontiers in neuroscience}, vol.~8, 2014.

\bibitem{gok2016prediction}
S.~Gok and M.~Sahin, ``Prediction of forelimb muscle emgs from the
  corticospinal signals in rats,'' in \emph{Engineering in Medicine and Biology
  Society (EMBC), 2016 IEEE 38th Annual International Conference of the}.\hskip
  1em plus 0.5em minus 0.4em\relax IEEE, 2016, pp. 2780--2783.

\bibitem{kennedy1998restoration}
P.~R. Kennedy and R.~A. Bakay, ``Restoration of neural output from a paralyzed
  patient by a direct brain connection,'' \emph{Neuroreport}, vol.~9, no.~8,
  pp. 1707--1711, 1998.

\bibitem{kennedy2000direct}
P.~R. Kennedy \emph{et~al.}, ``Direct control of a computer from the human
  central nervous system,'' \emph{IEEE Transactions on rehabilitation
  engineering}, vol.~8, no.~2, pp. 198--202, 2000.

\bibitem{musallam2004cognitive}
S.~Musallam \emph{et~al.}, ``Cognitive control signals for neural
  prosthetics,'' \emph{Science}, vol. 305, no. 5681, pp. 258--262, 2004.

\bibitem{hochberg2006neuronal}
L.~R. Hochberg \emph{et~al.}, ``Neuronal ensemble control of prosthetic devices
  by a human with tetraplegia,'' \emph{Nature}, vol. 442, no. 7099, pp.
  164--171, 2006.

\bibitem{simeral2011neural}
J.~Simeral \emph{et~al.}, ``Neural control of cursor trajectory and click by a
  human with tetraplegia 1000 days after implant of an intracortical
  microelectrode array,'' \emph{Journal of neural engineering}, vol.~8, no.~2,
  p. 025027, 2011.

\bibitem{hochberg2012reach}
L.~R. Hochberg \emph{et~al.}, ``Reach and grasp by people with tetraplegia
  using a neurally controlled robotic arm,'' \emph{Nature}, vol. 485, no. 7398,
  pp. 372--375, 2012.

\bibitem{collinger2013high}
J.~L. Collinger \emph{et~al.}, ``High-performance neuroprosthetic control by an
  individual with tetraplegia,'' \emph{The Lancet}, vol. 381, no. 9866, pp.
  557--564, 2013.

\bibitem{velliste2008cortical}
M.~Velliste \emph{et~al.}, ``Cortical control of a prosthetic arm for
  self-feeding,'' \emph{Nature}, vol. 453, no. 7198, pp. 1098--1101, 2008.

\bibitem{soekadar2016hybrid}
S.~Soekadar \emph{et~al.}, ``Hybrid eeg/eog-based brain/neural hand exoskeleton
  restores fully independent daily living activities after quadriplegia,''
  \emph{Science Robotics}, vol.~1, no.~1, p. eaag3296, 2016.

\bibitem{moritz2008direct}
C.~T. Moritz \emph{et~al.}, ``Direct control of paralysed muscles by cortical
  neurons,'' \emph{Nature}, vol. 456, no. 7222, pp. 639--642, 2008.

\bibitem{pohlmeyer2009toward}
E.~A. Pohlmeyer \emph{et~al.}, ``Toward the restoration of hand use to a
  paralyzed monkey: brain-controlled functional electrical stimulation of
  forearm muscles,'' \emph{PloS one}, vol.~4, no.~6, p. e5924, 2009.

\bibitem{ethier2012restoration}
C.~Ethier \emph{et~al.}, ``Restoration of grasp following paralysis through
  brain-controlled stimulation of muscles,'' \emph{Nature}, vol. 485, no. 7398,
  pp. 368--371, 2012.

\bibitem{zimmermann2014closed}
J.~B. Zimmermann and A.~Jackson, ``Closed-loop control of spinal cord
  stimulation to restore hand function after paralysis,'' \emph{Frontiers in
  neuroscience}, vol.~8, p.~87, 2014.

\bibitem{capogrosso2016brain}
M.~Capogrosso \emph{et~al.}, ``A brain--spine interface alleviating gait
  deficits after spinal cord injury in primates,'' \emph{Nature}, vol. 539, no.
  7628, pp. 284--288, 2016.

\bibitem{bouton2016restoring}
C.~E. Bouton \emph{et~al.}, ``Restoring cortical control of functional movement
  in a human with quadriplegia,'' \emph{Nature}, vol. 533, no. 7602, pp.
  247--250, 2016.

\bibitem{harkema2011effect}
S.~Harkema \emph{et~al.}, ``Effect of epidural stimulation of the lumbosacral
  spinal cord on voluntary movement, standing, and assisted stepping after
  motor complete paraplegia: a case study,'' \emph{The Lancet}, vol. 377, no.
  9781, pp. 1938--1947, 2011.

\bibitem{donati2016long}
A.~R. Donati \emph{et~al.}, ``Long-term training with a brain-machine
  interface-based gait protocol induces partial neurological recovery in
  paraplegic patients,'' \emph{Scientific reports}, vol.~6, 2016.

\bibitem{balevi2013physical}
E.~Balevi and O.~B. Akan, ``A physical channel model for nanoscale neuro-spike
  communications,'' \emph{IEEE Tran. Com.}, vol.~61, no.~3, pp. 1178--1187,
  2013.

\bibitem{malak2014communication}
D.~Malak and O.~Akan, ``Communication theoretical understanding of intra-body
  nervous nanonetworks,'' \emph{IEEE Communications Magazine}, vol.~52, no.~4,
  pp. 129--135, 2014.

\bibitem{malak2013communication}
D.~Malak and O.~Akan, ``A communication theoretical analysis of synaptic
  multiple-access channel in hippocampal-cortical neurons,'' \emph{IEEE Tran.
  Com.}, vol.~61, no.~6, pp. 2457--2467, 2013.

\bibitem{ramezani2017communication}
H.~Ramezani and O.~B. Akan, ``A communication theoretical modeling of axonal
  propagation in hippocampal pyramidal neurons,'' \emph{IEEE Transactions on
  NanoBioscience}, 2017.

\bibitem{Ramezani2015}
H.~Ramezani and O.~B. Akan, ``Synaptic channel model including effects of spike
  width variation,'' in \emph{2nd ACM NANOCOM}.\hskip 1em plus 0.5em minus
  0.4em\relax ACM, 2015.

\bibitem{Ramezani2017Importance}
H.~Ramezani and O.~B. Akan, ``Importance of vesicle release stochasticity in
  neuro-spike communication,'' in \emph{39th IEEE EMBC Conf.}\hskip 1em plus
  0.5em minus 0.4em\relax IEEE, 2017.

\bibitem{Ramezani2017Rate}
H.~Ramezani \emph{et~al.}, ``Rate region analysis of multi-terminal neuronal
  nanoscale communication channel,'' in \emph{17th IEEE NANO Conf.}\hskip 1em
  plus 0.5em minus 0.4em\relax IEEE, 2017.

\bibitem{ramezani2017information}
H.~Ramezani and O.~B. Akan, ``Information capacity of vesicle release in
  neuro-spike communication,'' \emph{IEEE Communications Letters}, 2017.

\bibitem{khan2017diffusion}
T.~Khan \emph{et~al.}, ``Diffusion-based model for synaptic molecular
  communication channel,'' \emph{IEEE Transactions on NanoBioscience}, 2017.

\bibitem{abbasi2015queueing}
N.~A. Abbasi and O.~B. Akan, ``A queueing-theoretical delay analysis for
  intra-body nervous nanonetwork,'' \emph{Nano Com. Net.}, 2015.

\bibitem{abbasi2018controlled}
N.~A. Abbasi \emph{et~al.}, ``Controlled information transfer through an in
  vivo nervous system,'' \emph{Scientific reports}, vol.~8, no.~1, p. 2298,
  2018.

\bibitem{purves2004neuroscience}
D.~Purves \emph{et~al.}, ``Neuroscience. 3rd,'' \emph{Massachusetts: Sinauer
  Associates Inc Publishers}, 2004.

\bibitem{civas2020rate}
M.~Civas and O.~B. Akan, ``Rate of information flow across layered neuro-spike
  network in the spinal cord,'' \emph{IEEE Transactions on NanoBioscience},
  2020.

\bibitem{lemon2008descending}
R.~N. Lemon, ``Descending pathways in motor control,'' \emph{Annu. Rev.
  Neurosci.}, vol.~31, pp. 195--218, 2008.

\bibitem{McDonald_2002}
J.~W. McDonald and C.~Sadowsky, ``Spinal-cord injury,'' \emph{The Lancet}, vol.
  359, p. 417–425, Feb 2002.

\bibitem{Silva_2014}
S.~N. A. \emph{et~al.}, ``From basics to clinical: A comprehensive review on
  spinal cord injury,'' \emph{Progress in Neurobiology}, vol. 114, p. 25–57,
  2014.

\bibitem{11_papastefanakimatsas_2015}
R.~Papastefanaki, FlorentiaMatsas, ``From demyelination to remyelination : The
  road toward therapies for spinal cord injury,'' \emph{Glia}, vol.~63, no.~7,
  pp. 1101--1125, 2015.

\bibitem{Oyinbo_2011}
C.~A. Oyinbo, ``Secondary injury mechanisms in traumatic spinal cord injury: a
  nugget of this multiply cascade.'' \emph{Spinal Cord}, vol.~71, pp. 281--299,
  2011.

\bibitem{5_kabu_2015}
S.~Kabu \emph{et~al.}, ``Drug delivery, cell-based therapies, and tissue
  engineering approaches for spinal cord injury,'' \emph{Journal of Controlled
  Release}, vol. 219, pp. 141--154, 2015.

\bibitem{llobet2019axon}
A.~Llobet~Rosell and L.~J. Neukomm, ``Axon death signalling in wallerian
  degeneration among species and in disease,'' \emph{Open biology}, vol.~9,
  no.~8, p. 190118, 2019.

\bibitem{prasad2012can}
A.~Prasad and M.~Sahin, ``Can motor volition be extracted from the spinal
  cord?'' \emph{Journal of neuroengineering and rehabilitation}, vol.~9, no.~1,
  p.~41, 2012.

\bibitem{kim2017spinal}
Y.-H. Kim \emph{et~al.}, ``Spinal cord injury and related clinical trials,''
  \emph{Clinics in orthopedic surgery}, vol.~9, no.~1, pp. 1--9, 2017.

\bibitem{tyler2013nanomedicine}
J.~Y. Tyler \emph{et~al.}, ``Nanomedicine for treating spinal cord injury,''
  \emph{Nanoscale}, vol.~5, no.~19, pp. 8821--8836, 2013.

\bibitem{faccendini2017nanofiber}
A.~Faccendini \emph{et~al.}, ``Nanofiber scaffolds as drug delivery systems to
  bridge spinal cord injury,'' \emph{Pharmaceuticals}, vol.~10, no.~3, p.~63,
  2017.

\bibitem{willerth2007approaches}
S.~M. Willerth and S.~E. Sakiyama-Elbert, ``Approaches to neural tissue
  engineering using scaffolds for drug delivery,'' \emph{Advanced drug delivery
  reviews}, vol.~59, no.~4, pp. 325--338, 2007.

\bibitem{lu2012long}
P.~Lu \emph{et~al.}, ``Long-distance growth and connectivity of neural stem
  cells after severe spinal cord injury,'' \emph{Cell}, vol. 150, no.~6, pp.
  1264--1273, 2012.

\bibitem{nakajima2012transplantation}
H.~Nakajima \emph{et~al.}, ``Transplantation of mesenchymal stem cells promotes
  an alternative pathway of macrophage activation and functional recovery after
  spinal cord injury,'' \emph{Journal of neurotrauma}, vol.~29, no.~8, pp.
  1614--1625, 2012.

\bibitem{sharp2010human}
J.~Sharp \emph{et~al.}, ``Human embryonic stem cell-derived oligodendrocyte
  progenitor cell transplants improve recovery after cervical spinal cord
  injury,'' \emph{Stem cells}, vol.~28, no.~1, pp. 152--163, 2010.

\bibitem{kanno2014combination}
H.~Kanno \emph{et~al.}, ``Combination of engineered schwann cell grafts to
  secrete neurotrophin and chondroitinase promotes axonal regeneration and
  locomotion after spinal cord injury,'' \emph{Journal of Neuroscience},
  vol.~34, no.~5, pp. 1838--1855, 2014.

\bibitem{11_pearse_2007}
D.~D. Pearse \emph{et~al.}, ``Transplantation of schwann cells and/or olfactory
  ensheathing glia into the contused spinal cord: Survival, migration, axon
  association, and functional recovery,'' \emph{Glia}, vol.~55, no.~9, pp.
  976--1000, 2007.

\bibitem{krsko2009length}
P.~Krsko \emph{et~al.}, ``Length-scale mediated adhesion and directed growth of
  neural cells by surface-patterned poly (ethylene glycol) hydrogels,''
  \emph{Biomaterials}, vol.~30, no.~5, pp. 721--729, 2009.

\bibitem{mahoney2006three}
M.~J. Mahoney and K.~S. Anseth, ``Three-dimensional growth and function of
  neural tissue in degradable polyethylene glycol hydrogels,''
  \emph{Biomaterials}, vol.~27, no.~10, pp. 2265--2274, 2006.

\bibitem{liu2013self}
Y.~Liu \emph{et~al.}, ``A self-assembling peptide reduces glial scarring,
  attenuates post-traumatic inflammation and promotes neurological recovery
  following spinal cord injury,'' \emph{Acta biomaterialia}, vol.~9, no.~9, pp.
  8075--8088, 2013.

\bibitem{guo2007reknitting}
J.~Guo \emph{et~al.}, ``Reknitting the injured spinal cord by self-assembling
  peptide nanofiber scaffold,'' \emph{Nanomedicine: Nanotechnology, Biology and
  Medicine}, vol.~3, no.~4, pp. 311--321, 2007.

\bibitem{tysseling2008self}
V.~M. Tysseling-Mattiace \emph{et~al.}, ``Self-assembling nanofibers inhibit
  glial scar formation and promote axon elongation after spinal cord injury,''
  \emph{Journal of Neuroscience}, vol.~28, no.~14, pp. 3814--3823, 2008.

\bibitem{chen2015repair}
B.~Chen \emph{et~al.}, ``Repair of spinal cord injury by implantation of
  bfgf-incorporated hema-moetacl hydrogel in rats,'' \emph{Scientific Reports
  (Nature Publisher Group)}, vol.~5, p. 9017, 2015.

\bibitem{milbreta2016three}
U.~Milbreta \emph{et~al.}, ``Three-dimensional nanofiber hybrid scaffold
  directs and enhances axonal regeneration after spinal cord injury,''
  \emph{ACS Biomaterials Science \& Engineering}, vol.~2, no.~8, pp.
  1319--1329, 2016.

\bibitem{nguyen2017three}
L.~H. Nguyen \emph{et~al.}, ``Three-dimensional aligned nanofibers-hydrogel
  scaffold for controlled non-viral drug/gene delivery to direct axon
  regeneration in spinal cord injury treatment,'' \emph{Scientific Reports},
  vol.~7, p. 42212, 2017.

\bibitem{gautam2017engineering}
V.~Gautam \emph{et~al.}, ``Engineering highly interconnected neuronal networks
  on nanowire scaffolds,'' \emph{Nano letters}, vol.~17, no.~6, pp. 3369--3375,
  2017.

\bibitem{timashev20163d}
P.~Timashev \emph{et~al.}, ``3d in vitro platform produced by two-photon
  polymerization for the analysis of neural network formation and function,''
  \emph{Biomedical Physics \& Engineering Express}, vol.~2, no.~3, p. 035001,
  2016.

\bibitem{demarse2016feed}
T.~B. DeMarse \emph{et~al.}, ``Feed-forward propagation of temporal and rate
  information between cortical populations during coherent activation in
  engineered in vitro networks,'' \emph{Frontiers in neural circuits}, vol.~10,
  p.~32, 2016.

\bibitem{gladkov2017design}
A.~Gladkov \emph{et~al.}, ``Design of cultured neuron networks in vitro with
  predefined connectivity using asymmetric microfluidic channels,''
  \emph{Scientific Reports}, vol.~7, no.~1, p. 15625, 2017.

\bibitem{assinck2017cell}
P.~Assinck \emph{et~al.}, ``Cell transplantation therapy for spinal cord
  injury,'' \emph{Nature Neuroscience}, vol.~20, no.~05, pp. 637--647, 2017.

\bibitem{vismara2017current}
I.~Vismara \emph{et~al.}, ``Current options for cell therapy in spinal cord
  injury,'' \emph{Trends in Molecular Medicine}, vol.~23, no.~9, pp. 831--849,
  2017.

\bibitem{chen2016advancing}
F.-M. Chen and X.~Liu, ``Advancing biomaterials of human origin for tissue
  engineering,'' \emph{Progress in polymer science}, vol.~53, pp. 86--168,
  2016.

\bibitem{rajangam2016wireless}
S.~Rajangam \emph{et~al.}, ``Wireless cortical brain-machine interface for
  whole-body navigation in primates,'' \emph{Scientific reports}, vol.~6, p.
  22170, 2016.

\bibitem{boraud2002single}
T.~Boraud \emph{et~al.}, ``From single extracellular unit recording in
  experimental and human parkinsonism to the development of a functional
  concept of the role played by the basal ganglia in motor control,''
  \emph{Progress in neurobiology}, vol.~66, no.~4, pp. 265--283, 2002.

\bibitem{wolpaw2002brain}
J.~R. Wolpaw \emph{et~al.}, ``Brain--computer interfaces for communication and
  control,'' \emph{Clinical neurophysiology}, vol. 113, no.~6, pp. 767--791,
  2002.

\bibitem{gangulyintroduction}
K.~Ganguly, ``An introduction to brain-machine interfaces.''

\bibitem{cincotti2008high}
F.~Cincotti \emph{et~al.}, ``High-resolution eeg techniques for brain--computer
  interface applications,'' \emph{Journal of neuroscience methods}, vol. 167,
  no.~1, pp. 31--42, 2008.

\bibitem{craig2007adaptive}
D.~A. Craig and H.~Nguyen, ``Adaptive eeg thought pattern classifier for
  advanced wheelchair control,'' in \emph{2007 29th Annual International
  Conference of the IEEE Engineering in Medicine and Biology Society}.\hskip
  1em plus 0.5em minus 0.4em\relax IEEE, 2007, pp. 2544--2547.

\bibitem{leuthardt2006electrocorticography}
E.~C. Leuthardt \emph{et~al.}, ``Electrocorticography-based brain computer
  interface-the seattle experience,'' \emph{IEEE Transactions on Neural Systems
  and Rehabilitation Engineering}, vol.~14, no.~2, pp. 194--198, 2006.

\bibitem{leuthardt2004brain}
E.~C. Leuthardt \emph{et~al.}, ``A brain--computer interface using
  electrocorticographic signals in humans,'' \emph{Journal of neural
  engineering}, vol.~1, no.~2, p.~63, 2004.

\bibitem{schalk2007decoding}
G.~Schalk \emph{et~al.}, ``Decoding two-dimensional movement trajectories using
  electrocorticographic signals in humans,'' \emph{Journal of neural
  engineering}, vol.~4, no.~3, p. 264, 2007.

\bibitem{spuler2014decoding}
M.~Sp{\"u}ler \emph{et~al.}, ``Decoding of motor intentions from epidural ecog
  recordings in severely paralyzed chronic stroke patients,'' \emph{Journal of
  neural engineering}, vol.~11, no.~6, p. 066008, 2014.

\bibitem{wang2013electrocorticographic}
W.~Wang \emph{et~al.}, ``An electrocorticographic brain interface in an
  individual with tetraplegia,'' \emph{PloS one}, vol.~8, no.~2, p. e55344,
  2013.

\bibitem{hotson2016individual}
G.~Hotson \emph{et~al.}, ``Individual finger control of a modular prosthetic
  limb using high-density electrocorticography in a human subject,''
  \emph{Journal of neural engineering}, vol.~13, no.~2, p. 026017, 2016.

\bibitem{carmena2003learning}
J.~M. Carmena \emph{et~al.}, ``Learning to control a brain--machine interface
  for reaching and grasping by primates,'' \emph{PLoS biol}, vol.~1, no.~2, p.
  e42, 2003.

\bibitem{flint2013long}
R.~D. Flint \emph{et~al.}, ``Long term, stable brain machine interface
  performance using local field potentials and multiunit spikes,''
  \emph{Journal of neural engineering}, vol.~10, no.~5, p. 056005, 2013.

\bibitem{so2014subject}
K.~So \emph{et~al.}, ``Subject-specific modulation of local field potential
  spectral power during brain--machine interface control in primates,''
  \emph{Journal of neural engineering}, vol.~11, no.~2, p. 026002, 2014.

\bibitem{hill2012recording}
N.~J. Hill \emph{et~al.}, ``Recording human electrocorticographic (ecog)
  signals for neuroscientific research and real-time functional cortical
  mapping,'' \emph{JoVE (Journal of Visualized Experiments)}, no.~64, pp.
  e3993--e3993, 2012.

\bibitem{schalk2011brain}
G.~Schalk and E.~C. Leuthardt, ``Brain-computer interfaces using
  electrocorticographic signals,'' \emph{IEEE reviews in biomedical
  engineering}, vol.~4, pp. 140--154, 2011.

\bibitem{wang2009human}
W.~Wang \emph{et~al.}, ``Human motor cortical activity recorded with micro-ecog
  electrodes, during individual finger movements,'' in \emph{Engineering in
  Medicine and Biology Society, 2009. EMBC 2009. Annual International
  Conference of the IEEE}.\hskip 1em plus 0.5em minus 0.4em\relax IEEE, 2009,
  pp. 586--589.

\bibitem{chao2010long}
Z.~C. Chao \emph{et~al.}, ``Long-term asynchronous decoding of arm motion using
  electrocorticographic signals in monkey,'' \emph{Frontiers in
  neuroengineering}, vol.~3, p.~3, 2010.

\bibitem{cheung2007implantable}
K.~C. Cheung, ``Implantable microscale neural interfaces,'' \emph{Biomedical
  microdevices}, vol.~9, no.~6, pp. 923--938, 2007.

\bibitem{gunasekera2015intracortical}
B.~Gunasekera \emph{et~al.}, ``Intracortical recording interfaces: current
  challenges to chronic recording function,'' \emph{ACS chemical neuroscience},
  vol.~6, no.~1, pp. 68--83, 2015.

\bibitem{krusienski2006evaluation}
D.~J. Krusienski \emph{et~al.}, ``An evaluation of autoregressive spectral
  estimation model order for brain-computer interface applications,'' in
  \emph{2006 International Conference of the IEEE Engineering in Medicine and
  Biology Society}.\hskip 1em plus 0.5em minus 0.4em\relax IEEE, 2006, pp.
  1323--1326.

\bibitem{cleophas2013machine}
T.~J. Cleophas \emph{et~al.}, \emph{Machine learning in medicine}.\hskip 1em
  plus 0.5em minus 0.4em\relax Springer, 2013.

\bibitem{alpaydin2004introduction}
E.~Alpaydin, ``Introduction to machine learning (adaptive computation and
  machine learning series),'' \emph{The MIT Press Cambridge}, 2004.

\bibitem{biship2007pattern}
C.~M. Biship, ``Pattern recognition and machine learning (information science
  and statistics),'' 2007.

\bibitem{brunner2010improved}
C.~Brunner \emph{et~al.}, ``Improved signal processing approaches in an offline
  simulation of a hybrid brain--computer interface,'' \emph{Journal of
  neuroscience methods}, vol. 188, no.~1, pp. 165--173, 2010.

\bibitem{marsland2011machine}
S.~Marsland, \emph{Machine learning: an algorithmic perspective}.\hskip 1em
  plus 0.5em minus 0.4em\relax Chapman and Hall/CRC, 2011.

\bibitem{nicolas2012brain}
L.~F. Nicolas-Alonso and J.~Gomez-Gil, ``Brain computer interfaces, a review,''
  \emph{Sensors}, vol.~12, no.~2, pp. 1211--1279, 2012.

\bibitem{barber2012bayesian}
D.~Barber, \emph{Bayesian reasoning and machine learning}.\hskip 1em plus 0.5em
  minus 0.4em\relax Cambridge University Press, 2012.

\bibitem{goodfellow2016deep}
I.~Goodfellow \emph{et~al.}, \emph{Deep learning}.\hskip 1em plus 0.5em minus
  0.4em\relax MIT press, 2016.

\bibitem{lecun2015deep}
Y.~LeCun \emph{et~al.}, ``Deep learning,'' \emph{Nature}, vol. 521, no. 7553,
  pp. 436--444, 2015.

\bibitem{dube2014fatigue}
J.~Dube, ``Fatigue resistibility and stimulus strength using intraspinal
  microstimulation vs. intramuscular stimulation in a rat model: Case study,''
  2014.

\bibitem{grahn2015wireless}
P.~J. Grahn \emph{et~al.}, ``Wireless control of intraspinal microstimulation
  in a rodent model of paralysis,'' \emph{Journal of neurosurgery}, vol. 123,
  no.~1, pp. 232--242, 2015.

\bibitem{deisseroth2011optogenetics}
K.~Deisseroth, ``Optogenetics,'' \emph{Nature methods}, vol.~8, no.~1, p.~26,
  2011.

\bibitem{courtine2009transformation}
G.~Courtine \emph{et~al.}, ``Transformation of nonfunctional spinal circuits
  into functional states after the loss of brain input,'' \emph{Nature
  neuroscience}, vol.~12, no.~10, pp. 1333--1342, 2009.

\bibitem{wenger2014closed}
N.~Wenger \emph{et~al.}, ``Closed-loop neuromodulation of spinal sensorimotor
  circuits controls refined locomotion after complete spinal cord injury,''
  \emph{Science translational medicine}, vol.~6, no. 255, pp.
  255ra133--255ra133, 2014.

\bibitem{bamford2010effects}
J.~A. Bamford \emph{et~al.}, ``The effects of intraspinal microstimulation on
  spinal cord tissue in the rat,'' \emph{Biomaterials}, vol.~31, no.~21, pp.
  5552--5563, 2010.

\bibitem{caggiano2014rostro}
V.~Caggiano \emph{et~al.}, ``Rostro-caudal inhibition of hindlimb movements in
  the spinal cord of mice,'' \emph{PloS one}, vol.~9, no.~6, p. e100865, 2014.

\bibitem{caggiano2016optogenetic}
V.~Caggiano \emph{et~al.}, ``An optogenetic demonstration of motor modularity
  in the mammalian spinal cord,'' \emph{Scientific reports}, vol.~6, p. 35185,
  2016.

\bibitem{montgomery2016beyond}
K.~L. Montgomery \emph{et~al.}, ``Beyond the brain: Optogenetic control in the
  spinal cord and peripheral nervous system,'' \emph{Science translational
  medicine}, vol.~8, no. 337, pp. 337rv5--337rv5, 2016.

\bibitem{park2015soft}
S.~I. Park \emph{et~al.}, ``Soft, stretchable, fully implantable miniaturized
  optoelectronic systems for wireless optogenetics,'' \emph{Nature
  biotechnology}, vol.~33, no.~12, p. 1280, 2015.

\bibitem{montgomery2015wirelessly}
K.~L. Montgomery \emph{et~al.}, ``Wirelessly powered, fully internal
  optogenetics for brain, spinal and peripheral circuits in mice,''
  \emph{Nature methods}, vol.~12, no.~10, pp. 969--974, 2015.

\bibitem{weber2016miniaturized}
M.~J. Weber \emph{et~al.}, ``A miniaturized ultrasonically powered programmable
  optogenetic implant stimulator system,'' in \emph{Biomedical Wireless
  Technologies, Networks, and Sensing Systems (BioWireleSS), 2016 IEEE Topical
  Conference on}.\hskip 1em plus 0.5em minus 0.4em\relax IEEE, 2016, pp.
  12--14.

\bibitem{arbabian2016sound}
A.~Arbabian \emph{et~al.}, ``Sound technologies, sound bodies: Medical implants
  with ultrasonic links,'' \emph{IEEE Microwave Magazine}, vol.~17, no.~12, pp.
  39--54, 2016.

\bibitem{pfingst1977response}
B.~Pfingst \emph{et~al.}, ``Response plasticity of neurons in auditory cortex
  of the rhesus monkey,'' \emph{Experimental Brain Research}, vol.~29, no. 3-4,
  pp. 393--404, 1977.

\bibitem{larsson2013organic}
K.~C. Larsson \emph{et~al.}, ``Organic bioelectronics for
  electronic-to-chemical translation in modulation of neuronal signaling and
  machine-to-brain interfacing,'' \emph{Biochimica et Biophysica Acta
  (BBA)-General Subjects}, vol. 1830, no.~9, pp. 4334--4344, 2013.

\bibitem{beeby2006energy}
S.~P. Beeby \emph{et~al.}, ``Energy harvesting vibration sources for
  microsystems applications,'' \emph{Measurement science and technology},
  vol.~17, no.~12, p. R175, 2006.

\bibitem{pisarchik2019novel}
A.~N. Pisarchik \emph{et~al.}, ``From novel technology to novel applications:
  Comment on “an integrated brain-machine interface platform with thousands
  of channels” by elon musk and neuralink,'' \emph{Journal of medical
  Internet research}, vol.~21, no.~10, p. e16356, 2019.

\bibitem{mead1990neuromorphic}
C.~Mead, ``Neuromorphic electronic systems,'' \emph{Proceedings of the IEEE},
  vol.~78, no.~10, pp. 1629--1636, 1990.

\bibitem{saighi2015plasticity}
S.~Sa{\"\i}ghi \emph{et~al.}, ``Plasticity in memristive devices for spiking
  neural networks,'' \emph{Frontiers in neuroscience}, vol.~9, p.~51, 2015.

\bibitem{otero2012biomimetic}
T.~Otero \emph{et~al.}, ``Biomimetic electrochemistry from conducting polymers.
  a review: artificial muscles, smart membranes, smart drug delivery and
  computer/neuron interfaces,'' \emph{Electrochimica Acta}, vol.~84, pp.
  112--128, 2012.

\bibitem{simon2015organic}
D.~T. Simon \emph{et~al.}, ``An organic electronic biomimetic neuron enables
  auto-regulated neuromodulation,'' \emph{Biosensors and Bioelectronics},
  vol.~71, pp. 359--364, 2015.

\bibitem{ghez1990roles}
C.~Ghez \emph{et~al.}, ``Roles of proprioceptive input in the programming of
  arm trajectories,'' in \emph{Cold spring harbor symposia on quantitative
  biology}, vol.~55.\hskip 1em plus 0.5em minus 0.4em\relax Cold Spring Harbor
  Laboratory Press, 1990, pp. 837--847.

\bibitem{hatsopoulos2009science}
N.~G. Hatsopoulos and J.~P. Donoghue, ``The science of neural interface
  systems,'' \emph{Annual review of neuroscience}, vol.~32, pp. 249--266, 2009.

\bibitem{ramos2012proprioceptive}
A.~Ramos-Murguialday \emph{et~al.}, ``Proprioceptive feedback and brain
  computer interface (bci) based neuroprostheses,'' \emph{PloS one}, vol.~7,
  no.~10, p. e47048, 2012.

\bibitem{dhillon2005direct}
G.~S. Dhillon and K.~W. Horch, ``Direct neural sensory feedback and control of
  a prosthetic arm,'' \emph{IEEE transactions on neural systems and
  rehabilitation engineering}, vol.~13, no.~4, pp. 468--472, 2005.

\bibitem{horch2011object}
K.~Horch \emph{et~al.}, ``Object discrimination with an artificial hand using
  electrical stimulation of peripheral tactile and proprioceptive pathways with
  intrafascicular electrodes,'' \emph{IEEE Transactions on Neural Systems and
  Rehabilitation Engineering}, vol.~19, no.~5, pp. 483--489, 2011.

\bibitem{tyler2002functionally}
D.~J. Tyler and D.~M. Durand, ``Functionally selective peripheral nerve
  stimulation with a flat interface nerve electrode,'' \emph{IEEE Transactions
  on Neural Systems and Rehabilitation Engineering}, vol.~10, no.~4, pp.
  294--303, 2002.

\bibitem{rollings2016ion}
R.~C. Rollings \emph{et~al.}, ``Ion selectivity of graphene nanopores,''
  \emph{Nature communications}, vol.~7, 2016.

\bibitem{elias2009control}
D.~C. Elias \emph{et~al.}, ``Control of graphene's properties by reversible
  hydrogenation: evidence for graphane,'' \emph{Science}, vol. 323, no. 5914,
  pp. 610--613, 2009.

\bibitem{nair2010fluorographene}
R.~R. Nair \emph{et~al.}, ``Fluorographene: A two-dimensional counterpart of
  teflon,'' \emph{small}, vol.~6, no.~24, pp. 2877--2884, 2010.

\bibitem{li2011promotion}
N.~Li \emph{et~al.}, ``The promotion of neurite sprouting and outgrowth of
  mouse hippocampal cells in culture by graphene substrates,''
  \emph{Biomaterials}, vol.~32, no.~35, pp. 9374--9382, 2011.

\bibitem{fabbro2016graphene}
A.~Fabbro \emph{et~al.}, ``Graphene-based interfaces do not alter target nerve
  cells,'' \emph{ACS nano}, vol.~10, no.~1, pp. 615--623, 2016.

\bibitem{kuzum2014transparent}
D.~Kuzum \emph{et~al.}, ``Transparent and flexible low noise graphene
  electrodes for simultaneous electrophysiology and neuroimaging,''
  \emph{Nature communications}, vol.~5, no.~1, pp. 1--10, 2014.

\bibitem{blaschke2017mapping}
B.~Blaschke \emph{et~al.}, ``Mapping brain activity with flexible graphene
  micro-transistors. 2d mater,'' 2017.

\bibitem{lu2018graphene}
Y.~Lu \emph{et~al.}, ``Graphene-based neurotechnologies for advanced neural
  interfaces,'' \emph{Current Opinion in Biomedical Engineering}, vol.~6, pp.
  138--147, 2018.

\bibitem{lu2016flexible}
Y.~Lu \emph{et~al.}, ``Flexible neural electrode array based-on porous graphene
  for cortical microstimulation and sensing,'' \emph{Scientific reports},
  vol.~6, p. 33526, 2016.

\bibitem{hess2013graphene}
L.~H. Hess \emph{et~al.}, ``Graphene transistors for bioelectronics,''
  \emph{Proceedings of the IEEE}, vol. 101, no.~7, pp. 1780--1792, 2013.

\bibitem{huang2010nanoelectronic}
Y.~Huang \emph{et~al.}, ``Nanoelectronic biosensors based on cvd grown
  graphene,'' \emph{Nanoscale}, vol.~2, no.~8, pp. 1485--1488, 2010.

\bibitem{zhang_2016}
J.~Zhang \emph{et~al.}, ``Communication channel analysis and real time
  compressed sensing for high density neural recording devices,'' \emph{IEEE
  Transactions on Circuits and Systems}, vol.~63, no.~5, pp. 599--608, 2016.

\bibitem{kim2007thermal}
S.~Kim \emph{et~al.}, ``Thermal impact of an active 3-d microelectrode array
  implanted in the brain,'' \emph{IEEE Transactions on Neural Systems and
  Rehabilitation Engineering}, vol.~15, no.~4, pp. 493--501, 2007.

\bibitem{6572893}
K.~Abdelhalim \emph{et~al.}, ``64-channel uwb wireless neural vector analyzer
  soc with a closed-loop phase synchrony-triggered neurostimulator,''
  \emph{IEEE Journal of Solid-State Circuits}, vol.~48, no.~10, pp. 2494--2510,
  Oct 2013.

\bibitem{5201317}
O.~Novak \emph{et~al.}, ``Wireless ultra-wide-band data link for biomedical
  implants,'' in \emph{2009 Ph.D. Research in Microelectronics and
  Electronics}, July 2009, pp. 352--355.

\bibitem{6594741}
K.~M.~S. Thotahewa \emph{et~al.}, ``Power efficient ultra wide band based
  wireless body area networks with narrowband feedback path,'' \emph{IEEE
  Transactions on Mobile Computing}, vol.~13, no.~8, pp. 1829--1842, Aug 2014.

\bibitem{7469363}
J.~Zhang \emph{et~al.}, ``Communication channel analysis and real time
  compressed sensing for high density neural recording devices,'' \emph{IEEE
  Transactions on Circuits and Systems I: Regular Papers}, vol.~63, no.~5, pp.
  599--608, May 2016.

\bibitem{kim_2006}
S.~Kim \emph{et~al.}, ``Preliminary study of the thermal impact of a
  microelectrode array implanted in the brain,'' in \emph{Proceedings of the
  28th IEEE EMBS Annual International Conference}.\hskip 1em plus 0.5em minus
  0.4em\relax IEEE, Aug. 2006.

\bibitem{ture2018area}
K.~Ture \emph{et~al.}, ``Area and power efficient ultra-wideband transmitter
  based on active inductor,'' \emph{IEEE Transactions on Circuits and Systems
  II: Express Briefs}, vol.~65, no.~10, pp. 1325--1329, 2018.

\bibitem{tokgoz2018120gb}
K.~K. Tokgoz \emph{et~al.}, ``A 120gb/s 16qam cmos millimeter-wave wireless
  transceiver,'' in \emph{2018 IEEE International Solid-State Circuits
  Conference-(ISSCC)}.\hskip 1em plus 0.5em minus 0.4em\relax IEEE, 2018, pp.
  168--170.

\bibitem{7420751}
H.~Ando \emph{et~al.}, ``Wireless multichannel neural recording with a 128-mbps
  uwb transmitter for an implantable brain-machine interfaces,'' \emph{IEEE
  Transactions on Biomedical Circuits and Systems}, vol.~10, no.~6, pp.
  1068--1078, Dec 2016.

\bibitem{kabiri2017graphene}
S.~Kabiri~Ameri \emph{et~al.}, ``Graphene electronic tattoo sensors,''
  \emph{ACS nano}, vol.~11, no.~8, pp. 7634--7641, 2017.

\bibitem{lim2015transparent}
S.~Lim \emph{et~al.}, ``Transparent and stretchable interactive human machine
  interface based on patterned graphene heterostructures,'' \emph{Advanced
  Functional Materials}, vol.~25, no.~3, pp. 375--383, 2015.

\bibitem{kang2017graphene}
M.~Kang \emph{et~al.}, ``Graphene-based three-dimensional capacitive touch
  sensor for wearable electronics,'' \emph{ACS nano}, vol.~11, no.~8, pp.
  7950--7957, 2017.

\bibitem{zhang2016human}
Q.~Zhang \emph{et~al.}, ``Human-like sensing and reflexes of graphene-based
  films,'' \emph{Advanced Science}, vol.~3, no.~12, 2016.

\bibitem{kao2016duoskin}
H.-L.~C. Kao \emph{et~al.}, ``Duoskin: rapidly prototyping on-skin user
  interfaces using skin-friendly materials,'' in \emph{Proceedings of the 2016
  ACM International Symposium on Wearable Computers}.\hskip 1em plus 0.5em
  minus 0.4em\relax ACM, 2016, pp. 16--23.

\bibitem{huang2015binder}
X.~Huang \emph{et~al.}, ``Binder-free highly conductive graphene laminate for
  low cost printed radio frequency applications,'' \emph{Applied Physics
  Letters}, vol. 106, no.~20, p. 203105, 2015.

\bibitem{seo2016wireless}
D.~Seo \emph{et~al.}, ``Wireless recording in the peripheral nervous system
  with ultrasonic neural dust,'' \emph{Neuron}, vol.~91, no.~3, pp. 529--539,
  2016.

\bibitem{santagati2017experimental}
G.~E. Santagati and T.~Melodia, ``Experimental evaluation of impulsive
  ultrasonic intra-body communications for implantable biomedical devices,''
  \emph{IEEE Transactions on Mobile Computing}, vol.~16, no.~2, pp. 367--380,
  2017.

\bibitem{o2014selective}
S.~C. O’Hern \emph{et~al.}, ``Selective ionic transport through tunable
  subnanometer pores in single-layer graphene membranes,'' \emph{Nano letters},
  vol.~14, no.~3, pp. 1234--1241, 2014.

\bibitem{gerstner2002spiking}
W.~Gerstner and W.~M. Kistler, \emph{Spiking neuron models: Single neurons,
  populations, plasticity}.\hskip 1em plus 0.5em minus 0.4em\relax Cambridge
  university press, 2002.

\bibitem{li2017remarkable}
X.~Li \emph{et~al.}, ``Remarkable conductive anisotropy of metallic
  microcoil/pdms composites made by electric field induced alignment,''
  \emph{ACS applied materials \& interfaces}, vol.~9, no.~2, pp. 1593--1601,
  2017.

\bibitem{xu2015riemann}
F.~Xu \emph{et~al.}, ``Riemann surfaces of carbon as graphene nanosolenoids,''
  \emph{Nano letters}, vol.~16, no.~1, pp. 34--39, 2015.

\bibitem{tian2010design}
Y.~Tian \emph{et~al.}, ``Design and dynamics of a 3-dof flexure-based parallel
  mechanism for micro/nano manipulation,'' \emph{Microelectronic engineering},
  vol.~87, no.~2, pp. 230--241, 2010.

\bibitem{kralj2015magnetic}
S.~Kralj and D.~Makovec, ``Magnetic assembly of superparamagnetic iron oxide
  nanoparticle clusters into nanochains and nanobundles,'' \emph{ACS nano},
  vol.~9, no.~10, pp. 9700--9707, 2015.

\bibitem{khan2020nanosensor}
T.~Khan \emph{et~al.}, ``Nanosensor networks for smart health care,'' in
  \emph{Nanosensors for Smart Cities}.\hskip 1em plus 0.5em minus 0.4em\relax
  Elsevier, 2020, pp. 387--403.

\bibitem{akhtar2017energy}
F.~Akhtar and M.~H. Rehmani, ``Energy harvesting for self-sustainable wireless
  body area networks,'' \emph{IT Professional}, vol.~19, no.~2, pp. 32--40,
  2017.

\bibitem{yang2020challenges}
K.-W. Yang \emph{et~al.}, ``Challenges in scaling down of free-floating
  implantable neural interfaces to millimeter scale,'' \emph{IEEE Access},
  vol.~8, pp. 133\,295--133\,320, 2020.

\bibitem{akan2017internet}
O.~B. Akan \emph{et~al.}, ``Internet of hybrid energy harvesting things,''
  \emph{IEEE Internet of Things Journal}, vol.~5, no.~2, pp. 736--746, 2017.

\bibitem{wagih2020real}
M.~Wagih \emph{et~al.}, ``Real-world performance of sub-1 ghz and 2.4 ghz
  textile antennas for rf-powered body area networks,'' \emph{IEEE Access},
  vol.~8, pp. 133\,746--133\,756, 2020.

\bibitem{kuscu2018maximum}
M.~Kuscu and O.~B. Akan, ``Maximum likelihood detection with ligand receptors
  for diffusion-based molecular communications in internet of bio-nano
  things,'' \emph{IEEE transactions on nanobioscience}, vol.~17, no.~1, pp.
  44--54, 2018.

\bibitem{akyildiz2015internet}
I.~Akyildiz \emph{et~al.}, ``The internet of bio-nano things,'' \emph{IEEE Com.
  Mag.}, vol.~53, no.~3, pp. 32--40, 2015.

\bibitem{furber2016large}
S.~Furber, ``Large-scale neuromorphic computing systems,'' \emph{Journal of
  neural engineering}, vol.~13, no.~5, p. 051001, 2016.

\bibitem{cognolato2020gaze}
M.~Cognolato \emph{et~al.}, ``Gaze, visual, myoelectric, and inertial data of
  grasps for intelligent prosthetics,'' \emph{Scientific Data}, vol.~7, no.~1,
  pp. 1--15, 2020.

\bibitem{garry2019cyborgs}
T.~Garry and T.~Harwood, ``Cyborgs as frontline service employees: a research
  agenda,'' \emph{Journal of Service Theory and Practice}, 2019.

\end{thebibliography}
\end{document}